\documentclass[aps,pra,10pt,superscriptaddress,notitlepage,nofootinbib,twocolumn,floatfix]{revtex4-1}

\usepackage{amsmath}
\usepackage{graphicx}
\usepackage{color}
\usepackage[FIGTOPCAP,raggedright,nooneline]{subfigure}
\usepackage{amsfonts}
\usepackage{dsfont}
\newcommand{\dx}[2][]{\frac{\partial {#1}}{\partial #2}}
\newcommand{\ddx}[2][]{\frac{\partial^2 {#1}}{\partial #2^2}}
\newcommand{\abs}[1]{\left| {#1} \right|}

\newcommand{\E}{\mathcal{E}}
\newcommand{\dint}{\mathrm{d}}
\newcommand{\e}{\mathrm{e}}

\newcommand{\g}{\mathrm{g}}
\newcommand{\BO}{\mathrm{BO}}

\begin{document}
\title{Universality in a one-dimensional three-body system}

\author{Lucas Happ}
\email{lucas.happ@uni-ulm.de}
\affiliation{Institut f\"{u}r Quantenphysik and Center for Integrated Quantum Science and Technology (IQ$^{\rm ST}$),  Universit\"{a}t Ulm, D-89069 Ulm, Germany}
\author{Matthias Zimmermann}
\affiliation{Institut f\"{u}r Quantenphysik and Center for Integrated Quantum Science and Technology (IQ$^{\rm ST}$),  Universit\"{a}t Ulm, D-89069 Ulm, Germany}
\author{Santiago I. Betelu}
\affiliation{Department of Mathematics, University of North Texas, Denton, Texas 76203-5017, USA}
\author{Wolfgang P. Schleich}
\affiliation{Institut f\"{u}r Quantenphysik and Center for Integrated Quantum Science and Technology (IQ$^{\rm ST}$),  Universit\"{a}t Ulm, D-89069 Ulm, Germany}
\affiliation{Hagler Institute for Advanced Study, Institute for Quantum Science and Engineering (IQSE), and Texas A\&M AgriLife Research, Texas A\&M University, College Station, Texas 77843-4242, USA.}
\author{Maxim A. Efremov}
\affiliation{Institut f\"{u}r Quantenphysik and Center for Integrated Quantum Science and Technology (IQ$^{\rm ST}$),  Universit\"{a}t Ulm, D-89069 Ulm, Germany}

\date{\today}

\begin{abstract}
We study a heavy-heavy-light three-body system confined to one space dimension. Both binding energies and corresponding wave functions are obtained for (i) the zero-range, and (ii) two finite-range attractive heavy-light interaction potentials. In case of the zero-range potential, we apply the method of Skorniakov and Ter-Martirosian to explore the accuracy of the Born-Oppenheimer approach. For the finite-range potentials, we solve the Schr\"odinger equation numerically using a pseudospectral method. We demonstrate that when the two-body ground state energy approaches zero, the three-body bound states display a universal behavior, independent of the shape of the interaction potential.
\end{abstract}

\maketitle

\section{Introduction}
The few-body problem has been of central interest in the physics community since the very beginning of quantum mechanics \cite{BO1927,Heitler1927,Bethe1957,Richter1993}.
Continuous efforts have led to theoretical breakthroughs like the Efimov effect \cite{Efimov1970,*Efimov1971,*Efimov1973}, that is the appearance of an infinite sequence of universal bound states in the three-dimensional system of three bodies, provided the two-body interactions have a single $s$-wave resonance \cite{Nishida2012}. The effect is universal \cite{Nielsen2001,Jensen2004,Braaten2006,Greene2017,Naidon2017} in the sense that it is independent of the shape of the two-body interaction potential, as long as the latter is tuned to be on $s$-wave resonance.

In the present article we study another class of universal bound states in a three-body system of two identical, heavy particles and a third, light particle, all confined to one spatial dimension (1D) when the heavy-light ground state energy approaches zero. This nearly resonant state is not a virtual state but always weakly bound in the case of an attractive heavy-light interaction. We assume no interaction between the two heavy particles and obtain the binding energies as well as the corresponding wave functions for the zero- and two different finite-range heavy-light interaction potentials. In addition, we prove the universality of these states.

\subsection{Dimension of space and symmetry of resonance}
The appearance of the Efimov effect crucially depends on the number of spatial dimensions, and on the symmetry of the underlying two-body resonance. Indeed, changing in \textit{three dimensions} the symmetry of the two-body resonance from an $s$- to a $p$-\textit{wave} \cite{Efremov2013,Zhu2013} results in the reduction of the \textit{infinite} number of bound states to a \textit{finite} one.

Moreover, in the case of a \textit{two}- \cite{Bruch1979,Lim1980,Vugalter1983,Levinsen2014} or \textit{one-dimensional} \cite{Kartavtsev2009,Mehta2014} space, a two-body $s$-\textit{wave} resonance does not lead to the Efimov effect. Again the spectrum of the three-body bound states is {\it finite} and determined by the mass ratios between the particles \cite{Pricoupenko2010,Bellotti2013,Ngampruetikorn2013}. However, the {\it two-dimensional} system of three particles with a $p$\textit{-wave} inter-particle resonance can again support an {\it infinite} number of universal bound states, the so-called ``super Efimov'' effect \cite{Nishida2013,Moroz2014,Gridnev2014,Volosniev2014,Gao2015}.

Experimentally the changes in the number of space dimensions and the interaction can be implemented. Indeed, the reduction of the dimensionality is achieved by using off-resonant light to confine ultra-cold gases in quasi-1D or quasi-2D geometries \cite{Bloch2008}. In addition, the interactions between ultracold atoms can be tuned easily via Feshbach-resonances \cite{Chin2010}.

\subsection{Methods}
We solve the exact integral equations \cite{STM1957} of Skorniakov and Ter-Martirosian (STM) for the zero-range heavy-light interaction potential and obtain the three-body bound states for arbitrary mass ratios. Based on these exact results, we investigate the accuracy of the Born-Oppenheimer (BO) approximation \cite{BO1927} for the three-body problem depending on the mass ratio.

By considering finite-range potentials of Gaussian and cubic Lorentzian shape, we explore the universal regime. For these finite-range potentials we obtain the bound states of the three-particle system numerically using a pseudospectral method \cite{boyd2001chebyshev,trefethen2000spectral,BAYE20151} based on the roots of the rational Chebyshev functions.

\subsection{Overview}
Our article is organized as follows. In Section \ref{sec:statement} we briefly summarize the essential ingredients of the two- and three-body system. We then focus in Section \ref{sec:contact} on the case of the zero-range heavy-light interaction and utilize the BO approximation and the STM method. Next we dedicate Section \ref{sec:general_interaction} to a study of the universal behavior for two different finite-range potentials. In Section \ref{sec:proof} we then demonstrate the universality of the three-body bound states for any heavy-light interaction. We conclude in Section \ref{sec:summary} by summarizing our results and by presenting an outlook.

In order to keep our article self-contained but focused on the central ideas, we present more detailed calculations in two appendices. Appendix \ref{app:BO} is focused on the derivation of the diagonal correction to the BO approximation. In Appendix \ref{app:SM}, we introduce a grid based on the roots of the rational Chebyshev functions and recall briefly the pseudospectral method applied in Section \ref{sec:general_interaction}.

\section{The three-body system}\label{sec:statement}
In this section we first briefly discuss the validity of 1D models to describe quasi-1D systems. We then introduce the quantities determining an interacting mass-imbalanced two-body system in 1D. Next, we extend this system to the case of three interacting particles using dimensionless Jacobi coordinates. Finally, we discuss the corresponding Schr\"odinger equation and its symmetries, which serves as the basis for the studies presented in the subsequent sections.

\subsection{1D and quasi-1D models}\label{quasi1D}
Many theoretical studies \cite{Mehta2007,Kartavtsev2009,Mehta2014,Nishida2018,Guijarro2018} of three-body systems confined along two directions are performed using 1D models. This reduction offers the advantage of a simple and intuitive description revealing the underlying three-body properties. However, it is important to emphasize that experiments on these confined systems are always performed in quasi-1D.

In the case of a zero-range interaction the effective interaction potential of two particles in a tight cylindrical symmetric trap (quasi-1D setup) is given by the zero-range potential with the 1D scattering length determined by the 3D scattering length and the harmonic potential width, as shown in Ref. \cite{Olshanii1998}. Moreover, the dependence of universal three-body bound states on the dimensionality has been investigated in Refs. \cite{Levinsen2014,Zinner2015,Zinner2018,Pricoupenko2018}. In particular, when reducing the dimensionality from 3D to quasi-2D, the conditions to reproduce the results obtained by a 2D model are presented.

These results justify the relevance of 1D models for quasi-1D experiments and we thus analyze in the present article the three-body system using a 1D model.

\subsection{Two interacting particles}
\label{sec:two-body-interaction}
We consider a two-body system consisting of a heavy particle of mass~$M$ and a light one of mass~$m$, both constrained to 1D and interacting via a potential of range $\xi_0$.

After eliminating the heavy-light center-of-mass coordinate, the system is governed by the stationary Schr\"odinger equation 
\begin{equation}
\left[-\frac{1}{2}\frac{\textrm{d}^2}{\textrm{d}x^2}+v\left(x \right)\right]\psi^{(2)}=\mathcal{E}^{(2)}\psi^{(2)}
\label{eq: Schroedinger_2body_dmless}
\end{equation}
for the two-body wave function $\psi^{(2)}= \psi^{(2)}(x)$ of the relative motion presented in dimensionless units. Indeed, $x$ denotes the relative coordinate of the light particle with respect to the heavy one in units of the characteristic length $\xi_0$.

The two-body binding energy $\mathcal{E}^{(2)}$ and the  potential 
\begin{equation}\label{interaction}
v\left(x \right)= v_0 f(x)
\end{equation}
are both given in units of $ \hbar^2/\mu \xi_0^2$ with the Planck's constant $\hbar$ and the reduced mass $\mu\equiv Mm/(m+M)$ of the heavy-light system. Here $v_0$ denotes the magnitude and $f=f(x)$ the shape of the interaction potential.

We assume an attractive interaction, $v_0 < 0$, as well as a symmetric shape $f$, that is $f(x) = f(\abs{x})$. Moreover, we choose $v$ such that (i) it describes a short-range interaction, \textit{i.e.} $\abs{x}^2 f(|x|) \to 0$ as $\abs{x} \to \infty$, and (ii) the potential $v$ supports only a single bound state with energy $\mathcal{E}^{(2)}_\mathrm{g}$ and even wave function $\psi^{(2)}_\mathrm{g}(x)=\psi^{(2)}_\mathrm{g}(-x)$.

\subsection{Three interacting particles}
\label{sec:threebody1D}
We now add a third particle of mass $M$, also constrained to 1D and identical to the heavy particle in the heavy-light system considered above. Accordingly, we assume the same interaction potential~$v$ between the additional heavy and the light particle, but no interaction between the two heavy ones.

\begin{figure}[htbp]
\includegraphics[width =.8\columnwidth]{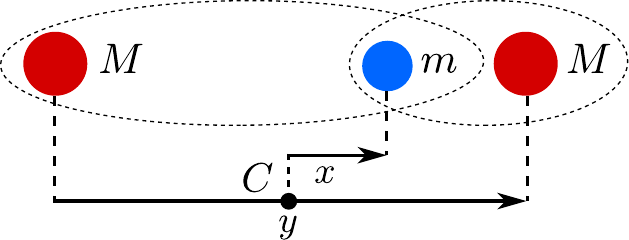}
\caption{Jacobi-coordinates $x$ and $y$ for the three particles confined to 1D.}
\label{fig:coords}
\end{figure}

Next, we introduce dimensionless Jacobi coordinates~\cite{Greene2017} $x$ and $y$ as displayed in Fig.~\ref{fig:coords}, where $y$ is the relative coordinate between the two heavy particles, and $x$ denotes the coordinate of the light particle with respect to the center-of-mass $C$ of the two heavy ones, both in units of $\xi_0$.

Eliminating again the center-of-mass motion of this heavy-heavy-light system, we arrive at the dimensionless stationary Schr\"odinger equation
\begin{equation}\label{TISGL}
\left[-\frac{\alpha_x}{2}\ddx{x} -\frac{\alpha_y}{2} \ddx{y} + v\left(r_+\right) + v\left(r_- \right) \right] \psi=  \mathcal{E} \psi
\end{equation}
for the three-body wave function $\psi = \psi(x,y)$ describing only the relative motions with $r_\pm \equiv x\pm y/2$.

The coefficients
\begin{equation}\label{eq: alphax}
\alpha_x \equiv \frac{1+2M/m}{2(1 +M/m)}
\end{equation}
and
\begin{equation}\label{eq: alphay}
\alpha_y \equiv \frac{2}{1 + M/m}
\end{equation}
depend only on the mass ratio $M/m$, and $\mathcal{E}$ denotes the dimensionless three-body energy in units of $ \hbar^2/\mu \xi_0^2$.

We notice that Eq. \eqref{TISGL} is invariant under the transformation $y\rightarrow -y$, that is an exchange of the two heavy particles. Hence, we distinguish even solutions $\psi(x,-y)=\psi(x,y)$ corresponding to two heavy {\it bosonic} particles, and odd solutions $\psi(x,-y)=-\psi(x,y)$ corresponding to two heavy {\it fermionic} particles. Moreover, also the transformation $x\rightarrow -x$ leaves Eq. \eqref{TISGL} invariant, and leads to the additional symmetry $\psi(-x,y) = \pm\psi(x,y)$.

\subsection{Formulation of the problem}
Now our aim is to solve Eq. (\ref{TISGL}) for the three-body bound state with the wave function $\psi_n$ and the corresponding energy $\E_n$ for $n=0,1,\ldots$. In particular, we are interested in the case when the two-body interaction described by the potential $v$ is close to a resonance, that is the energy $\E_\mathrm{g}^{(2)}$ of the two-body ground state approaches zero. Under the assumptions on $v$ presented in Sec. \ref{sec:two-body-interaction}, we then expect a universal behavior of the spectrum $\E_n$, namely that in the limit $\mathcal{E}_\mathrm{g}^{(2)}\to 0$ the ratio
\begin{equation}\label{epsilon}
\epsilon_n\equiv \frac{\mathcal{E}_n}{\left|\mathcal{E}_\mathrm{g}^{(2)}\right|}\,
\end{equation} 
is independent of the shape $f$ of the interaction potential.

In order to obtain a Hamiltonian with the eigenenergies $\epsilon_n$ given by Eq. (\ref{epsilon}), we introduce the rescaled variables
\begin{equation}\label{rescale1}
X \equiv x/\ell_\g, \qquad Y \equiv y/\ell_\g
\end{equation}
with
\begin{equation}\label{ell}
\ell_\g \equiv \frac{1}{\sqrt{2\abs{\E_\mathrm{g}^{(2)}}}},
\end{equation}
and rewrite Eq. (\ref{TISGL}) as the equation
\begin{equation}\label{rescale1b}
\hat{H} \tilde{\psi} = \epsilon \tilde{\psi}
\end{equation}
for the wave function $\tilde{\psi}=\tilde{\psi}(X,Y)$ with
\begin{equation}\label{rescale2}
\hat{H} \equiv -\alpha_x\ddx{X} - \alpha_y\ddx{Y} -2 \ell_\g^2 \abs{v_0} \bigg[f(\ell_\g R_+) + f(\ell_\g R_-)\bigg]
\end{equation} 
and 
\begin{equation}
R_\pm \equiv X \pm Y/2.
\end{equation}

We emphasize that the eigenvalues $\epsilon$ correspond to the ratio of the \textit{dimensional} three-body and two-body energies and are hence accessible in an experiment.

\section{Contact interaction}
\label{sec:contact}
We start our analysis by considering a contact interaction
\begin{equation}\label{eq: def_delta_potential}
f_\delta(x) \equiv \delta(x)
\end{equation}
between the light particle and each heavy one, with $\delta(x)$ being the Dirac delta function. For the two-body problem, this interaction potential $v_0 f_\delta(x)$ has only one bound state with the energy
\begin{equation}\label{e2bcontact}
\mathcal{E}_\mathrm{g}^{(2)} = -\frac{1}{2}v_0^{2}
\end{equation}
determined by the magnitude $v_0$ of the potential.

Using this relation and the scaling property of the Dirac delta function, $\delta(\alpha x) = \delta(x)/\abs{\alpha}$, we obtain
\begin{equation}\label{rescale3}
\ell_\g^2 \abs{v_0}  f_\delta\left(\ell_\g R_\pm\right) = \delta(X \pm Y/2),
\end{equation}
and the three-body Schr\"odinger equation, Eq. (\ref{rescale1b}), becomes independent of the interaction strength $v_0$. Hence $\epsilon$, the three-body binding energy in units of the two-body ground state energy, does not depend on $v_0$.

We now solve Eq. (\ref{rescale1b}) with $f=f_\delta$ using two different methods: the BO approximation \cite{BO1927} and an approach based on the exact STM integral equation \cite{STM1957}. We then compare the results of the two techniques to quantify the error of the BO approximation.

\subsection{Born-Oppenheimer approximation} \label{sec:BO}
The Born-Oppenheimer (BO) approach relies on approximating \cite{Efremov2009,Fonseca1979} the total three-body wave function $\tilde{\psi}$ in Eq. (\ref{rescale1b}) by the product
\begin{equation}\label{bo1}
\tilde{\psi}^{(\BO)}(X,Y) \equiv \varphi(X|Y)\phi(Y).
\end{equation}
Here, the wave function $\varphi(X|Y)$ describes the dynamics of the light particle in the potential of the two heavy ones, which are assumed to stay at a fixed distance $Y$.

The physical motivation of the ansatz Eq. (\ref{bo1}) is that for a large mass ratio, $M/m \gg 1$, the change of distance between the heavy particles is negligible on the relevant timescales of the light-particle dynamics. Hence, $Y$ does not change and enters in $\varphi$ only as a parameter, indicated by the vertical bar, giving rise to the Schr\"odinger equation
\begin{equation}\label{lpeq}
\left\{-\alpha_x \ddx{X} -2 \left[\delta(R_+) + \delta(R_-)\right] \right\}\varphi = u(Y) \varphi
\end{equation}
for the wave function $\varphi$ of the light particle, determining the so-called BO potential $u=u(Y)$.

In Appendix \ref{app:BOZO} we solve Eq. (\ref{lpeq}) analytically and obtain
\begin{equation}\label{BOpot}
u_\pm(Y) = -\frac{1}{\alpha_x}\left[\frac{\alpha_x}{\abs{Y}} W_0\left(\pm\frac{\abs{Y}}{\alpha_x}\e^{-\abs{Y}/\alpha_x}\right) +1 \right]^2
\end{equation}
expressed in terms of the Lambert function $W_0$ \cite{abramowitz}, and the corresponding wave functions
\begin{equation}\label{lpwf}
\varphi_\pm(X|Y) = N_\pm\left[ \e^{-\sqrt{\abs{u_\pm}/\alpha_x}\abs{R_-}} \pm \e^{-\sqrt{\abs{u_\pm}/\alpha_x}\abs{R_+}}\right],
\end{equation}
where $N_\pm$ is a normalization factor.

\begin{figure}[htbp]
\includegraphics[width=\columnwidth]{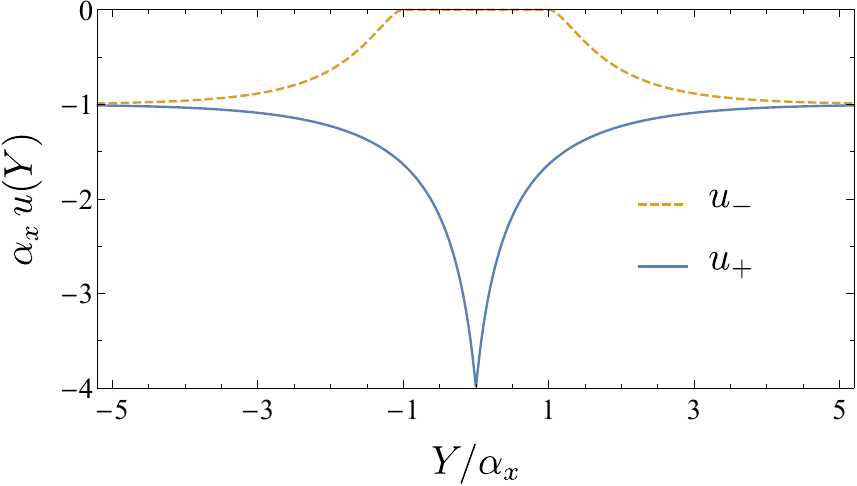}
\caption{Formation of three-body bound states explained by the two BO potentials $u_\pm=u_\pm(Y)$, Eq. (\ref{BOpot}), as a function of the relative distance $Y$ between the two heavy particles. Only the lower curve, corresponding to $u_+=u_+(Y)$, represents an attractive potential for the heavy particles and thus supports three-body bound states.}
\label{fig:BOPotential}
\end{figure}

The two potentials $u_\pm=u_\pm(Y)$ are displayed in Fig. \ref{fig:BOPotential}. Only the lower curve $u_+$,  corresponding to the symmetric light-particle state $\varphi_+$, provides an attractive potential for the two heavy particles and therefore supports bound states of the total three-body system.

The wave function $\phi_+=\phi_+(Y)$ of the heavy particles then obeys the Schr\"odinger equation
\begin{equation}\label{hpeq}
\left[ - \alpha_y \ddx{Y} + u_+(Y) \right] \phi_+ = \epsilon^{(\BO)} \phi_+,
\end{equation}
where $u_+ $ indeed plays the role of a potential, and $\epsilon^{(\BO)}$ is the scaled three-body energy within the BO approach.

Using the attractive potential $u_+$ given by Eq. (\ref{BOpot}), we calculate the values of $\epsilon_n^{(\BO)}$ numerically to a precision of $10^{-6}$ applying a pseudospectral method based on the Chebyshev grid introduced in Appendix \ref{app:SM}. It is important to mention that the scaled three-body bound state energies $\epsilon_n^{(\BO)}$ satisfy the inequality $\epsilon^{(\BO)} < -1/\alpha_x$, where the upper bound is the value of the BO potential at infinity,
\begin{equation}
\alpha_x u_+(Y\to\pm\infty) \to -1,
\end{equation}
as shown in Fig. \ref{fig:BOPotential}.

\begin{figure}[htbp]
\includegraphics[width=\columnwidth]{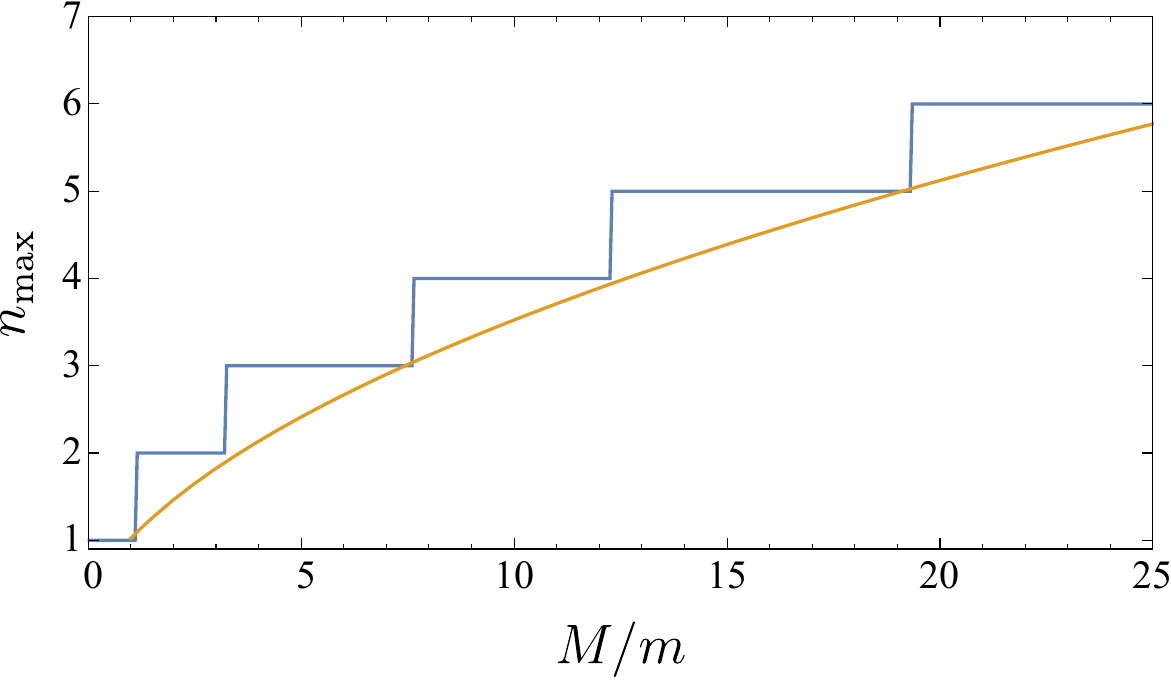}
\caption{Mass ratio $M/m$ determining the number $n_\mathrm{max}$ of three-body bound states (blue line) obtained numerically from Eq. (\ref{hpeq}) within the BO approximation together with the semiclassical lower bound (orange line), Eq. (\ref{nrs}). Increasing $M/m$ leads to more three-body bound states.}
\label{fig:Numberofstates}
\end{figure}

The number $n_\mathrm{max}$ of bound states supported by $u_+$ depends on the mass ratio $M/m$ and is depicted in Fig. \ref{fig:Numberofstates} as a blue line together with the semiclassical \cite{Schleich2001} estimation
\begin{equation}
n_\mathrm{max} \cong \frac{1}{\pi \sqrt{\alpha_y}} \int \dint Y\, \sqrt{\abs{u_+(Y)+\frac{1}{\alpha_x}}} -\frac{1}{2}
\end{equation}
or
\begin{equation}\label{nrs}
n_\mathrm{max} \cong 0.8781 \cdot \sqrt{1+2\frac{M}{m}} -\frac{1}{2}
\end{equation}
depicted by an orange line.

With increasing mass ratio $M/m$, additional bound states appear. A detailed comparison of the critical mass ratios required for the formation of a new bound state within the BO approximation and an hyperspherical approach can be found in Refs. \cite{Kartavtsev2009,Mehta2014}.

\subsection{Integral equation of Skorniakov and Ter-Martirosian}\label{sec:STM}
In this section we apply the method \cite{STM1957} of Skorniakov and Ter-Martirosian (STM) to the three-body problem described by Eqs. (\ref{rescale1b}) and (\ref{rescale2}) with a contact interaction. In contrast to the BO approach, this method does not involve any approximation, and in principle provides an exact solution for any mass ratio $M / m$.

We introduce the Green function
\begin{equation}\label{green2}
G_\epsilon^{(2)}(X,Y) \equiv -\frac{1}{2 \pi \sqrt{\alpha_x \alpha_y}} K_0\left(\sqrt{\abs{\epsilon}} \sqrt{\frac{1}{\alpha_x}X^2 + \frac{1}{\alpha_y}Y^2}\right)
\end{equation}
for the two-dimensional free-particle Schr\"odinger equation with $\epsilon <0$, where $K_0$ denotes the modified Bessel function of the second kind \cite{abramowitz}. We can then cast Eqs. (\ref{rescale1b}) and (\ref{rescale2}) in integral form
\begin{align}\label{stm1}
\tilde{\psi}(X,Y) = -&2\ell_\g^2 \abs{v_0} \iint \dint X'\dint Y' G_\epsilon^{(2)}(X-X',Y-Y') \nonumber \\
& \times \tilde{\psi}(X',Y') \left[f\left(\ell_\g R_+'\right) + f\left(\ell_\g R_-'\right)\right],
\end{align}
where $R_\pm' \equiv X' \pm Y'/2$.

With the help of Eq. (\ref{rescale3}) this expression simplifies in the special case of the contact interaction $f = f_\delta$ to
\begin{align}\label{stm2}
\tilde{\psi}(X&,Y) = -2 \iint \dint X'\dint Y' G_\epsilon^{(2)}(X-X',Y-Y') \nonumber \\
& \times \tilde{\psi}(X',Y') \left[\delta\left(X'+Y'/2\right) + \delta\left(X'-Y'/2\right)\right].
\end{align}

The delta functions then allow us to immediately perform the integration over $Y'$ and to obtain the one-dimensional integral equation
\begin{align}\label{stm3}
\tilde{\psi}(X,Y) = &-4 \int \dint X'\left[  G_\epsilon^{(2)}(X-X',Y-2X')\tilde{\psi}(X',2X') \right. \nonumber \\
+&\left.  G_\epsilon^{(2)}(X-X',Y+2X')\tilde{\psi}(X',-2X') \right].
\end{align}

\begin{figure*}[htbp]
\begin{center}
\subfigure[\hspace{\columnwidth}]{
\includegraphics[width=.85\columnwidth]{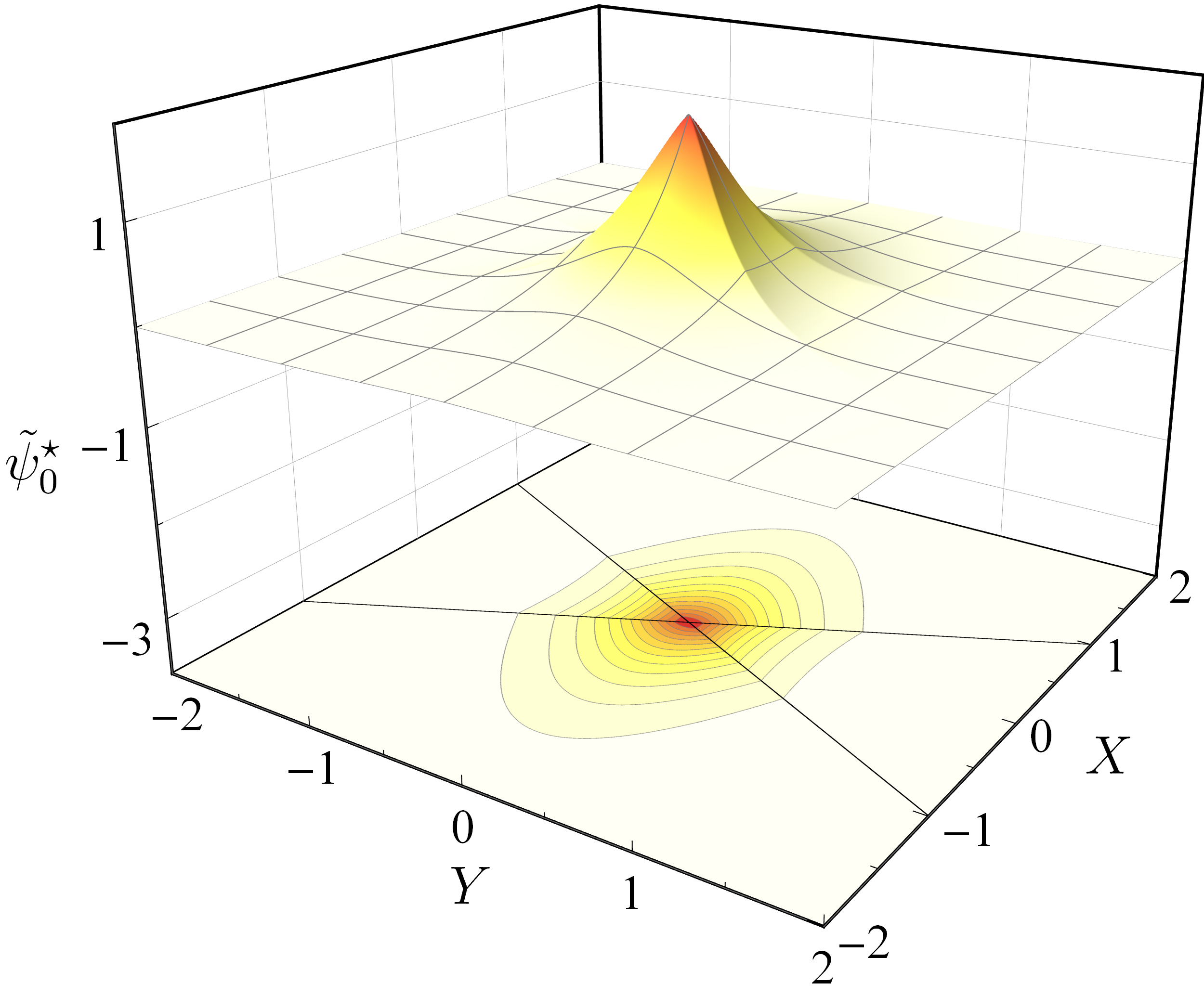}}
\hfill
\subfigure[\hspace{\columnwidth}]{
\includegraphics[width=.85\columnwidth]{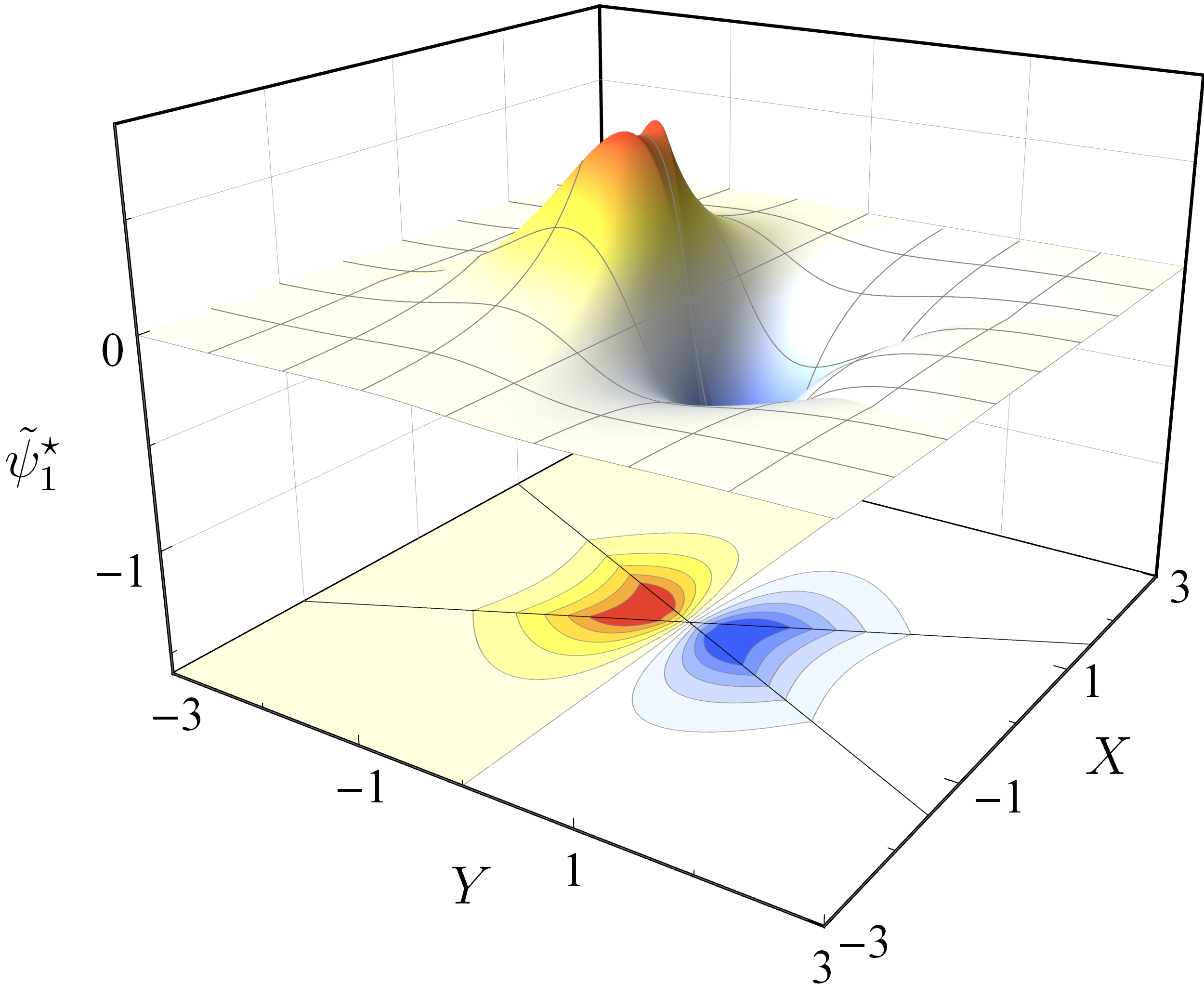}}\\
\subfigure[\hspace{\textwidth}]{\includegraphics[width=.85\columnwidth]{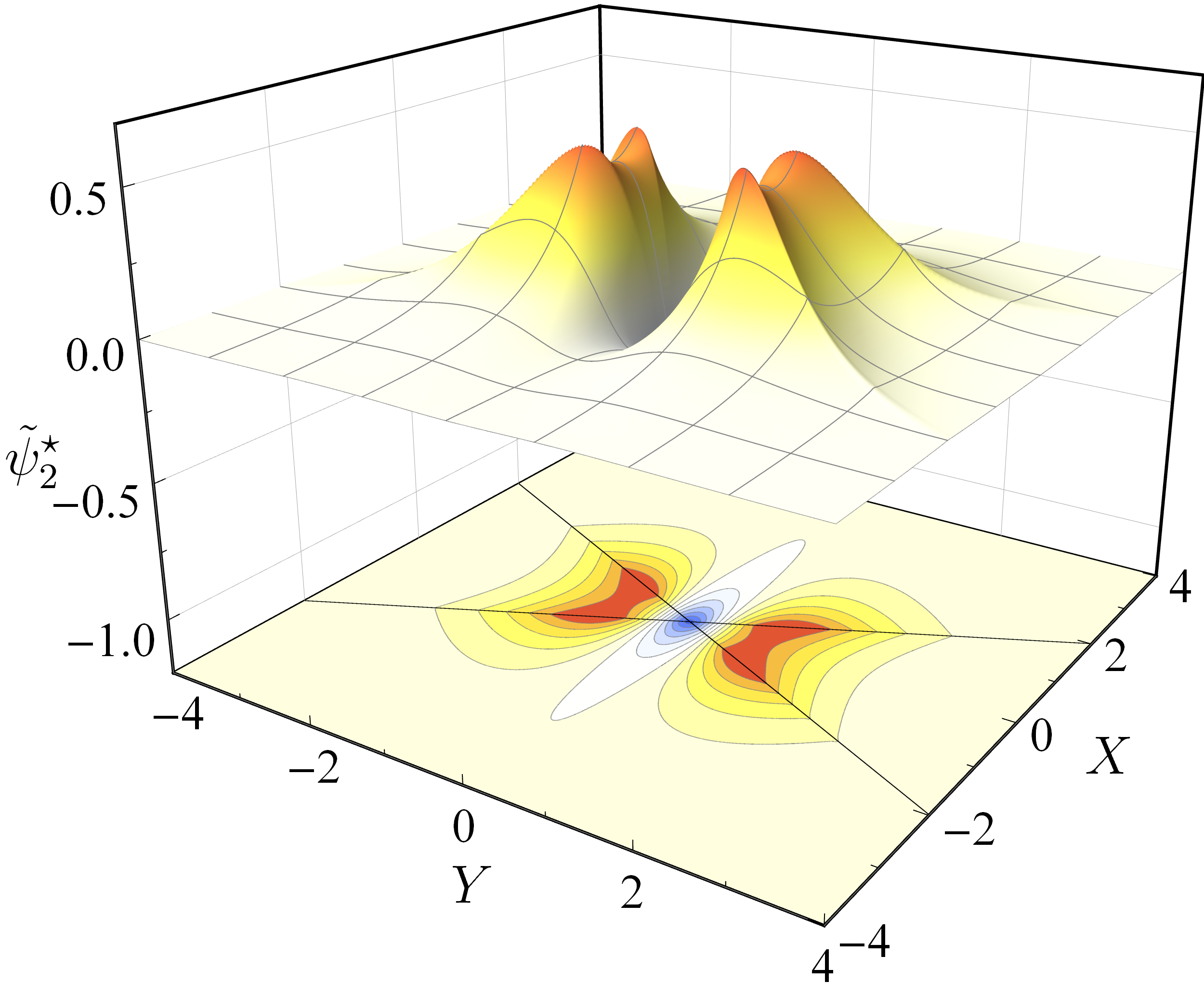}}
\hfill
\subfigure[\hspace{\textwidth}]{\includegraphics[width=.85\columnwidth]{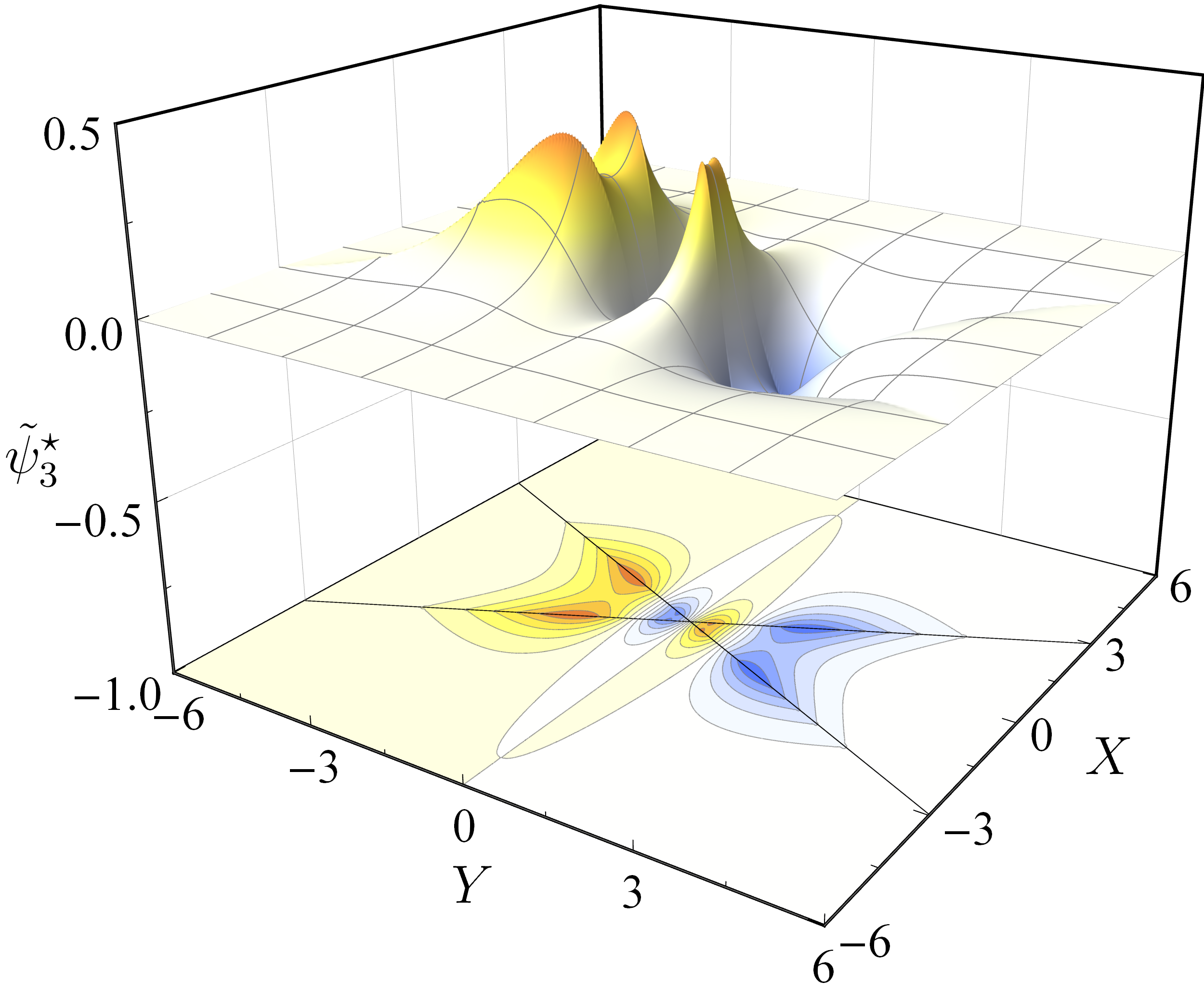}} \\
\end{center}
\caption{The lowest four (a)-(d) three-body wave functions $\tilde{\psi}_n^{\star}=\tilde{\psi}_n^{\star}(X,Y)$ corresponding to $n=0,1,2,3$ obtained by the integral equation of Skorniakov and Ter-Martirosian for a mass ratio $M/m = 20$, together with corresponding contour plots below. In addition, we depict the lines of interaction, that is $X \pm Y/2=0$, as black lines inside the contour plots. Due to the non-differentiability of the delta potential, the wave functions show a kink when crossing those lines perpendicularly. All states share the same symmetry in $X$-direction, whereas it alters from symmetric to antisymmetric in $Y$-direction. The transformation $Y\to -Y$ represents the exchange of the two heavy particles. Hence, the symmetry with respect to the line $Y=0$ indicates whether they are of \textit{bosonic} (symmetric case), or of \textit{fermionic} (antisymmetric case) character. The symmetry with respect to the line $X=0$ can be understood within the BO picture. Indeed, the lower-lying light-particle BO-wave function $\varphi_+=\varphi_+(X|Y)$ given by Eq. (\ref{lpwf}) leading to an attractive potential $u_+=u_+(Y)$ between the two heavy particles and thus to three-body bound states, is symmetric in $X$. Higher excited states have increased size, as indicated by the different scales in the plots.}
\label{fig:wf_plots}
\end{figure*}

Since the Hamiltonian $\hat{H}$, defined by  Eq. (\ref{rescale2}), is invariant under the transformation $Y \to - Y$, that is under the exchange of the two heavy particles, the solution $\tilde{\psi}$ has to be either even or odd with the symmetry relations
\begin{equation}\label{stm4}
\tilde{\psi}(X,-Y) = \pm \tilde{\psi}(X,Y)
\end{equation}
corresponding to the case of \textit{bosonic} (plus sign) or \textit{fermionic} (minus sign) heavy particles.

Evaluating both sides of Eq. (\ref{stm3}) at $Y = 2X$, we arrive at the integral equation
\begin{equation}\label{stm5}
\tilde{\psi}(X,2X) = \int \dint X'\, \mathcal{K}_\epsilon^{(\pm)}(X,X') \tilde{\psi}(X',2X')
\end{equation}
with the kernel
\begin{align}\label{kernel}
\mathcal{K}_\epsilon^{(\pm)}(X,X') \equiv -4&\left[G_\epsilon^{(2)}(X-X',2X-2X') \right. \nonumber \\
& \left. \pm G_\epsilon^{(2)}(X-X',2X+2X')\right],
\end{align}
where we have used the symmetry relation given by Eq. (\ref{stm4}) in writing $\tilde{\psi}(X',-2X') = \pm \tilde{\psi}(X',2X')$.

By rescaling the coordinates $X$ and $X'$ by $\sqrt{\abs{\epsilon}}$, Eq. (\ref{stm5}) can be cast into an eigenvalue problem for the eigenfunction $\tilde{\psi}(X/\sqrt{\abs{\epsilon}},2X/\sqrt{\abs{\epsilon}})$ with eigenvalue $\sqrt{\abs{\epsilon}}$, where the condition $\epsilon < -1$ determines the three-body bound states. Hence, the desired spectrum $\epsilon_n^\star$ of three-body bound states in units of the two-body ground state energy $\E_\mathrm{g}^{(2)}$ can be efficiently computed.

The three-body wave function $\tilde{\psi}(X,Y)$ is more difficult to obtain and requires an additional step. Together with the spectrum $\epsilon_n^\star$, we first obtain $\tilde{\psi}_n^\star(X,2X)$ from Eq. (\ref{stm5})
, that is $\tilde{\psi}_n^\star$ along the lines of interaction $Y = \pm 2X$. Then, we insert both $\epsilon_n^{\star}$ and $\tilde{\psi}_n^\star (X,2X)$ into the right-hand side of Eq. (\ref{stm3}). Taking into account the symmetry property $\tilde{\psi}_n^\star(X,-2X) = (-1)^{n}\tilde{\psi}_n^\star(X,2X)$ (even $n$ correspond to \textit{bosonic} heavy particles, whereas odd $n$ represent the \textit{fermionic} case) and performing the integration over $X'$, we finally obtain the entire three-body wave function $\tilde{\psi}_n^{\star}=\tilde{\psi}_n^\star (X,Y)$.

In Fig. \ref{fig:wf_plots} we depict the four ($n = 0,1,2,3$) lowest three-body bound states obtained via the STM method for $M/m = 20$. We emphasize again the scaling property of the delta function  yielding the wave function
\begin{equation}\label{wfscaling}
\psi_n^{\star}(x,y) = \tilde{\psi}_n\left(\sqrt{-2\E_\mathrm{g}^{(2)}}\,x,\sqrt{-2\E_\mathrm{g}^{(2)}}\,y\right)
\end{equation}
in the unscaled variables $x$ and $y$.

\subsection{BO approximation vs. STM approach}\label{stmvsbo}
In the preceding subsections we have applied the BO and STM methods to solve the 1D three-body problem with contact interaction. Now we compare the dependence of the resulting spectra and wave functions on the mass ratio $M/m$. Common experimental mass ratios range from $M/m=1$ for identical particles via $M / m \cong 2.2$ and $M / m \cong 12.4$ for $^{87}\mathrm{Rb}$--$^{40}\mathrm{K}$ and $^{87}\mathrm{Rb}$--$^{7}\mathrm{Li}$ mixtures respectively, to more extreme values of $M / m \cong 22.2$ in case of $^{133}\mathrm{Cs}$--$^{6}\mathrm{Li}$ mixtures \cite{Naidon2017}. Therefore we choose this range for our analysis.

\subsubsection{Energy spectrum}
\begin{figure}[htbp]
\subfigure[\hspace{\columnwidth}]{
\includegraphics[width=\columnwidth]{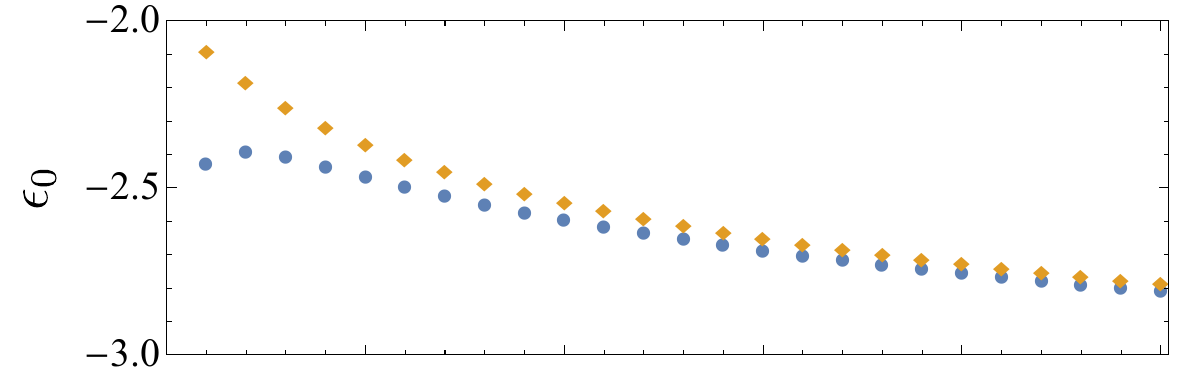}} \\
\subfigure[\hspace{\columnwidth}]{
\includegraphics[width=\columnwidth]{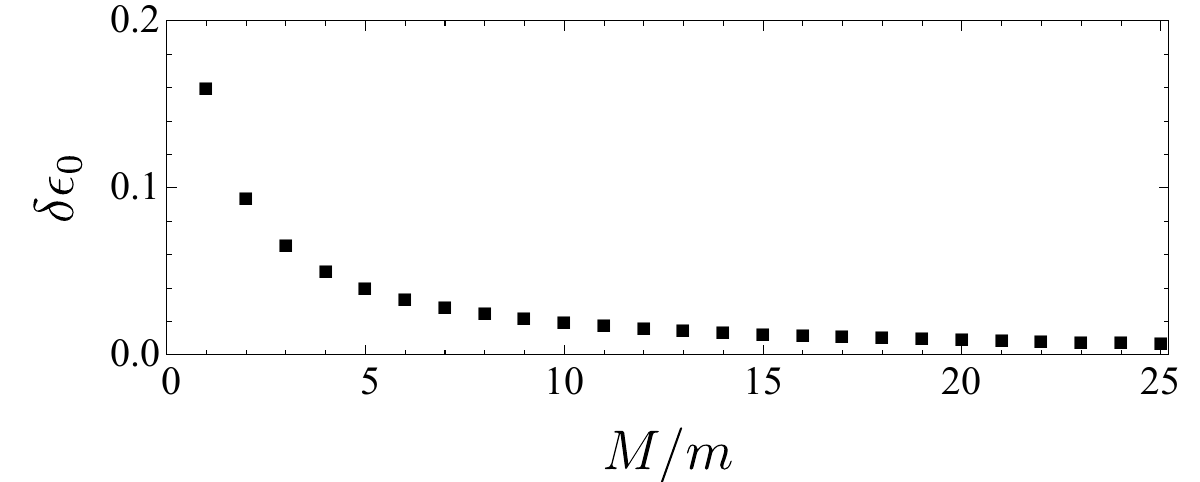}}
\caption{Accuracy of the BO ground state energies when compared to the corresponding STM values as a function of the mass ratio $M / m$. Here we depict (a) the energies $\epsilon_0$ of the lowest bound state, obtained via the approximate BO approach (blue dots), Eq. (\ref{hpeq}), and the exact STM integral equation (yellow diamonds), Eq. (\ref{stm5}). The BO approach underestimates the bound state energies, hence all blue dots lie below their yellow partners. With increasing mass ratio $M / m$ the blue dots approach the yellow ones, that is the BO approximation becomes more accurate for larger $M / m$. This behavior is further confirmed (b) by the monotonically decreasing relative error $\delta \epsilon_0$ (black squares), defined by Eq. (\ref{deviationE}).}
\label{fig:BO-STMSpectrum}
\end{figure}

In Fig. \ref{fig:BO-STMSpectrum} we display the three-body ground state energy $\epsilon_0$ obtained by the BO (blue dots) and the STM (yellow diamonds) method as a function of the mass ratio $M /m$. In addition, the relative error
\begin{equation}\label{deviationE}
\delta \epsilon_n \equiv (\epsilon_n^{(\BO)} - \epsilon_n^{\star})/\epsilon_n^{\star}
\end{equation}
is depicted as black squares for $n=0$.

\begin{figure*}[htb]
\centering
\subfigure[\hspace{\columnwidth}]{
\includegraphics[width=.45\textwidth]{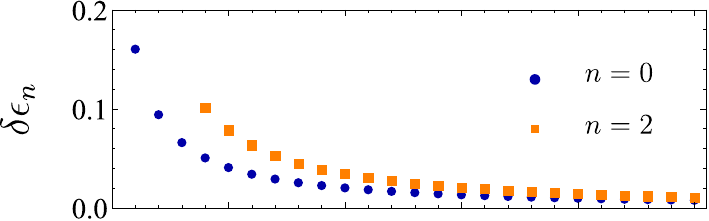}}
\hfill
\subfigure[\hspace{\columnwidth}]{
\includegraphics[width=.45\textwidth]{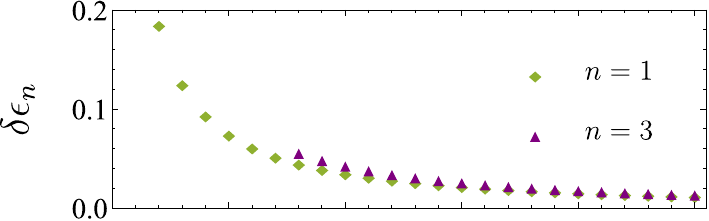}} \\
\subfigure[\hspace{\columnwidth}]{
\includegraphics[width=.448\textwidth]{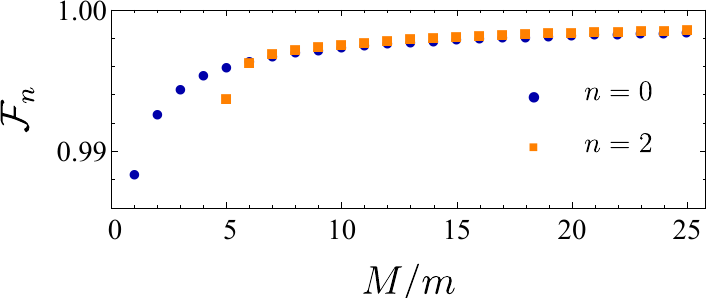}}
\hfill
\subfigure[\hspace{\columnwidth}]{
\includegraphics[width=.45\textwidth]{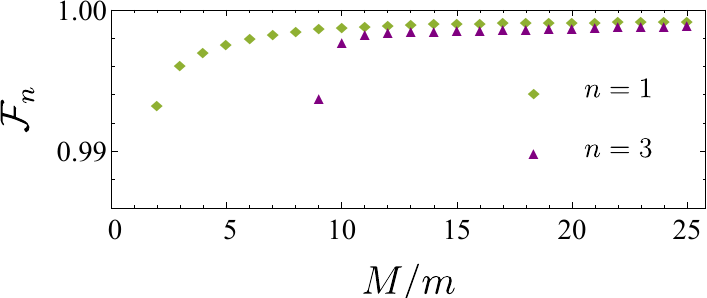}}
\caption{Comparison between BO approximation and exact STM approach based on the relative deviation in the three-body energy spectrum (top) together with fidelities between bound state wave functions (bottom). For the four lowest ($n = 0,1,2,3$) bound states, we display the relative deviation $\delta \epsilon_n$, Eq. (\ref{deviationE}), of the binding energies (top) and the fidelity $\mathcal{F}_n$, Eq. (\ref{fidelity}), of the corresponding bound state wave functions (bottom) as a function of the mass ratio $M / m$. We distinguish the cases of two \textit{bosonic} (a), (c) and two \textit{fermionic} (b), (d) heavy particles. The relative deviation $\delta \epsilon_n$ decreases monotonically from $16\%$ for $M/m=1$ to $2\%$ for $M/m = 25$. Higher excited states always show a higher deviation in the binding energy. For all states, the fidelity is increasing monotonically with increasing $M / m$, starting from 0.988 for $M/m = 1$ up to 0.999 for $M/m = 25$. However, no obvious dependence on $n$ is visible for $M/m \leq 25$. Fidelities corresponding to states very close to a resonance are not sufficiently converged and hence omitted in the picture.}
\label{fig:spectrum_fidelity}
\end{figure*}

For both methods the energies are computed numerically and are accurate up to $10^{-6}$. We find that our results are in excellent agreement with the values in the literature. Indeed, for $M/m=1$ we obtain $\epsilon_0^{(\BO)} = - 2.42267$ being within accuracy of the value $-2.4227$ in Ref. \cite{Mehta2014}, as well as $\epsilon_0^{\star} = - 2.087719$ which matches the value $-2.087719$ and is very close to $-2.08754$ found in Refs. \cite{Kartavtsev2009} and \cite{Gaudin1975}, respectively.

Likewise, we obtain the previously reported \cite{Mehta2014} error of about $16\,\%$ for the BO ground state energy $\epsilon_0^{(\BO)}$ for $M/m = 1$, a regime in which the BO approximation is not expected to provide reasonable results. Moreover, the relative error $\delta \epsilon_0$ decreases monotonically with increasing mass ratio $M / m$ and drops below $2\,\%$ for $M / m \cong 22.2$, as shown in Fig. \ref{fig:BO-STMSpectrum}.

We depict in the top row of Fig. \ref{fig:spectrum_fidelity} the relative errors $\delta \epsilon_n$ as a function of the mass ratio $M / m$ for the lowest four bound states $(n = 0,1,2,3)$. Excited states ($n\geq 1$) appear with increasing mass ratio, as shown in Fig. \ref{fig:Numberofstates}. The higher excited a state is, the larger the corresponding error gets. This behavior can be understood from the fact that the BO approximation involves neglecting derivatives, effecting more strongly higher excited states as they are more oscillatory.

\subsubsection{Wave functions}
After comparing the energy spectra calculated within the BO and the STM method, we now turn to the wave functions obtained by both methods and study the fidelity \cite{Nielsen2000} which for pure states simplifies to the spatial overlap
\begin{equation}\label{fidelity}
\mathcal{F}_n \equiv \left[\iint \dint X \dint Y\, \tilde{\psi}_n^{(\BO)}(X,Y)\tilde{\psi}_n^{\star}(X,Y)\right]^2
\end{equation}
of the wave functions $ \tilde{\psi}_n^{(\BO)}$ and $\tilde{\psi}_n^{\star}$. The BO wave function $\tilde{\psi}_n^{(\BO)}$ is a product of $\varphi_+$, Eq. (\ref{lpwf}), and $\phi_{+ n}$, obtained from Eq. (\ref{hpeq}), with $u_+$ given by Eq. (\ref{BOpot}), whereas the STM wave function $\tilde{\psi}_n^{\star}$, Eq. \eqref{wfscaling}, is calculated directly from Eq. (\ref{stm3}). Both functions are evaluated numerically on the same Chebyshev grid introduced in Appendix \ref{app:SM}, and the computed fidelities are accurate up to $10^{-4}$.

In the bottom row of Fig. \ref{fig:spectrum_fidelity} we present $\mathcal{F}_n$ for the four lowest $(n=0,1,2,3)$ bound states as a function of $M/m$. As expected, the fidelity increases monotonically for all bound states with increasing mass ratio $M/m$, that is the BO approximation becomes more accurate. However, it is remarkable that the fidelity $\mathcal{F}_n$ starts already from 0.988 at $M/m = 1$ and reaches values up to 0.999 for $M / m = 25$.

Moreover, $\mathcal{F}_n$ does not show a clear dependence on $n$: higher excited states do not always have a lower fidelity, in contrast to the expectation that the BO approximation should be worse for higher excited states. It is nevertheless possible that this behavior arises for much larger values of the ratio $M/m$.

\subsubsection{Diagonal energy correction}
We emphasize that the comparisons of the energy spectra and of the wave functions are based on different measures. The fidelity indicates how well a state can mimic another one in a measurement. As the fidelity is almost unity, we expect only a small deviation in the spectrum with respect to $\epsilon_n^{\star}$, if the exact state $\tilde{\psi}_n^{\star}$ is replaced by $\tilde{\psi}_n^{(\BO)}$ leading to the expression
\begin{equation}\label{ebo_1}
\bar{\epsilon}_n^{(\BO)} \equiv \iint \dint X \dint Y\, \tilde{\psi}_n^{(\BO)}(X,Y)\hat{H}\tilde{\psi}_n^{(\BO)}(X,Y)
\end{equation}
for the mean value of the energy. Here $\hat{H}$ is the full three-body Hamiltonian defined in Eq. (\ref{rescale2}) with $f=f_\delta$.

\begin{figure}[htbp]
\includegraphics[width=\columnwidth]{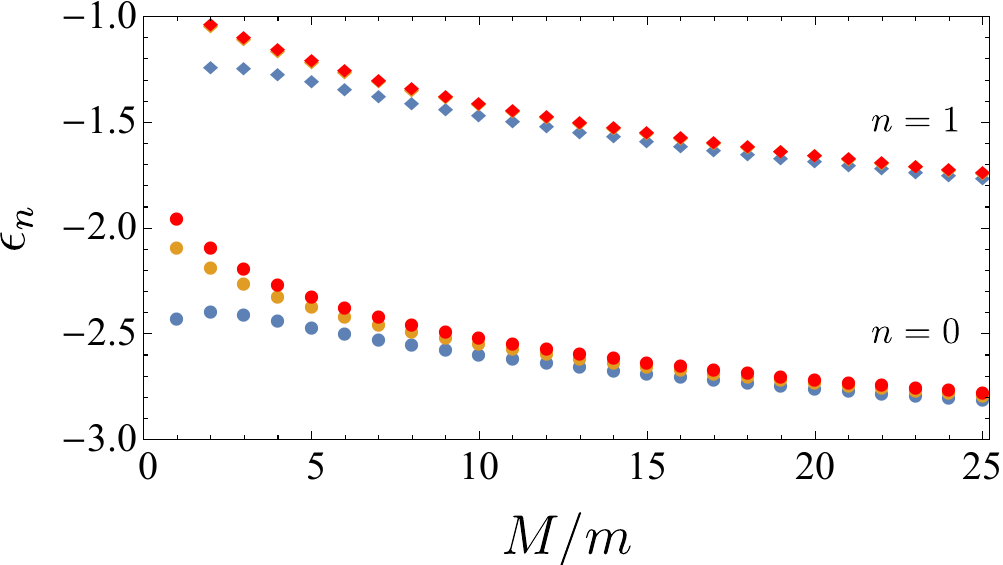}
\caption{Improved bound state energies $\bar{\epsilon}_n^{(\BO)}$ defined by Eq. (\ref{ebo_1}) (red symbols), $\epsilon_n^{(\BO)}$ (blue symbols) and $\epsilon_n^{\star}$ (yellow symbols) as a function of $M / m$. The energies are depicted by dots for the ground state ($n=0$), and by diamonds for the first ($n=1$) excited state. The deviation between $\bar{\epsilon}_n^{(\BO)}$ and $\epsilon_n^{\star}$ is reduced compared to the result in zero order, presented in Fig. \ref{fig:spectrum_fidelity}. For the first excited state, the red diamonds lie on top of the yellow ones, indicating that the values including the diagonal correction are almost indistinguishable from the exact ones.}
\label{fig:BO1}
\end{figure}

As shown in Appendix \ref{app:BOfirstorder}, $\bar{\epsilon}_n^{(\BO)}$ coincides with the BO energies including diagonal correction terms, and is depicted in Fig. \ref{fig:BO1} for the two lowest states ($n=0,1$). Compared to the zero-order BO approximation $\epsilon_n^{(\BO)}$, the deviation with respect to $\epsilon_n^{\star}$ is reduced by up to an order of magnitude. Hence, this feature suggests that the major contribution to the deviation between $\epsilon_n^{(\BO)}$ and $\epsilon_n^{\star}$ stems from the Hamiltonian itself, and not from the wave functions.

\subsubsection{Summary}
In summary, for the contact interaction, the BO approximation works surprisingly well in estimating the bound state energies, and even better for the corresponding wave functions. Moreover, for the contact interaction, the accuracy of the BO approximation is determined solely by the mass ratio between heavy and light particles and provides reasonable results even in the case of equal masses.

\section{General interaction potentials}
\label{sec:general_interaction}
So far we have only studied the case of a contact interaction between heavy and light particles. In this section we focus on different short-range interaction potentials and apply a pseudospectral method based on the roots of rational Chebyshev functions. 

In particular, we consider a two-body system close to a resonance and analyze the emergence of the universality in the mass-imbalanced three-body system. In this regime we retrieve for both the energy spectrum and the corresponding wave functions the results obtained for the case of a contact interaction. 

\subsection{Two-body interaction}
In this section we find numerically the relation between the two-body binding energy $\mathcal{E}^{(2)}_\g$ of the ground state, and the potential depth $v_0$ for different shapes $f$ of the interaction potential, Eq. (\ref{interaction}). For this purpose, we apply a pseudospectral method \cite{boyd2001chebyshev,trefethen2000spectral,BAYE20151} using a grid based on the roots of the rational Chebyshev functions \cite{BOYD1987112}.

According to Appendix \ref{app:SM}, we represent the dimensionless Schr\"odinger equation, Eq. \eqref{eq: Schroedinger_2body_dmless}, for the two-body system as a generalized eigenvalue problem 
\begin{equation}
\left[-\frac{1}{2}  \boldsymbol{D_2}- \mathcal{E}^{(2)}_\g\, \boldsymbol{\mathds{1}} \right]\vec{\psi}_\g^{{(2)}}=-v_0 \boldsymbol{F}\vec{\psi}_\g^{(2)}
\label{eq: two_body_discrete}
\end{equation}
with the generalized eigenvalue $-v_0$ and the generalized eigenvector $\vec{\psi}_\g^{(2)}$ of size $\mathcal{N}$ containing the values of the function $\psi_\g^{(2)}(x)$ evaluated at the grid points. The matrices $\boldsymbol{D_2}$ and $\boldsymbol{F}$ of size $\mathcal{N} \times \mathcal{N}$ are introduced in Appendix \ref{app: 1D_representation}, and  $\boldsymbol{\mathds{1}}$ denotes the identity matrix.

For a given two-body binding energy $\mathcal{E}^{(2)}_\g$ we determine the lowest generalized eigenvalue $-v_0=\left|v_0\right|$ of Eq. \eqref{eq: two_body_discrete} which specifies the potential depth, as well as the corresponding generalized eigenvector $\vec{\psi}_\g^{(2)}$, yielding an approximation to the wave function $\psi_\g^{(2)}(x)$ of the lowest state with the energy $\mathcal{E}^{(2)}_\g$.

We perform this calculation for two different interaction potentials $v=v(x)$, namely a potential 
\begin{equation}
f_\textrm{G}(x)\equiv\exp\left(-x^2\right)
\label{eq: def_Gaussian_potential}
\end{equation}
with a Gaussian shape and a potential
\begin{equation}
f_\textrm{L}(x)\equiv\frac{1}{\left(1+x^2\right)^3}
\label{eq: def_Lorentzian3_potential}
\end{equation}
being characterized by the cube of a Lorentzian.

\begin{figure}[htbp]
\includegraphics[width=\columnwidth]{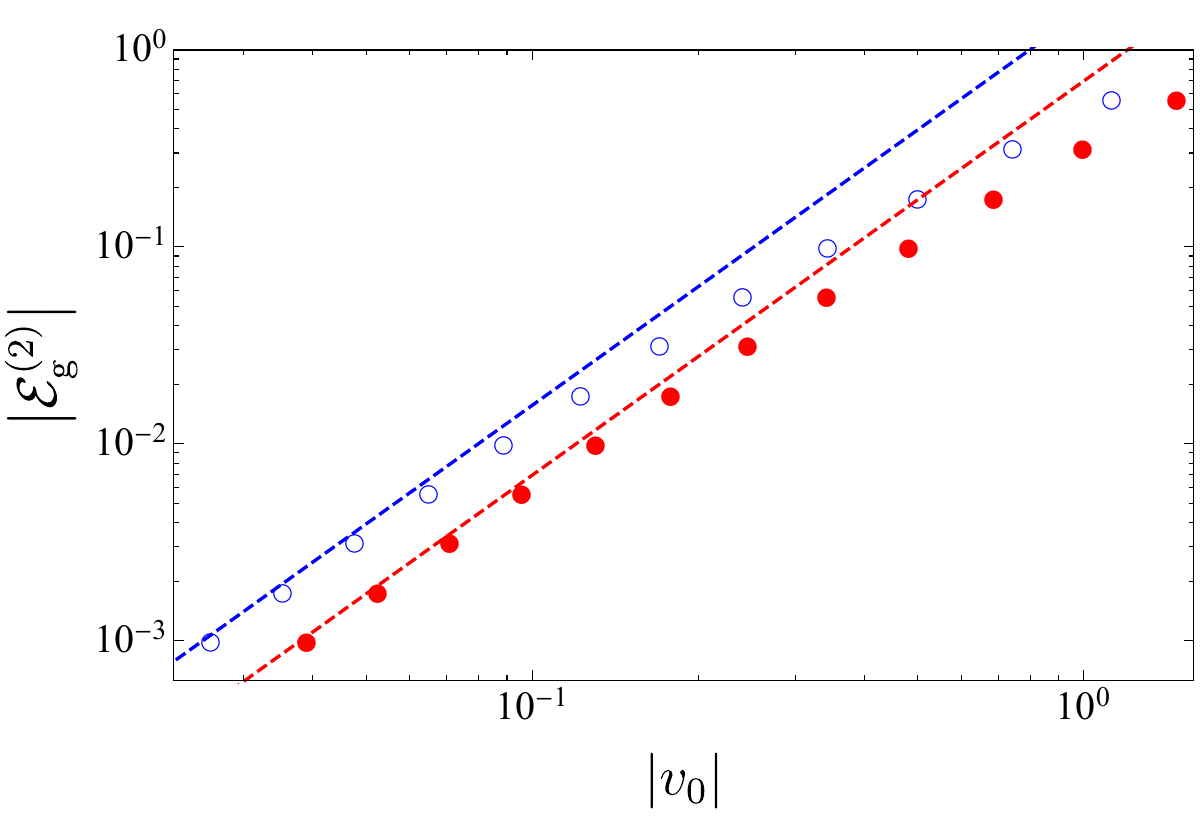}
\caption{Asymptotic behavior of the two-body ground state energy $\mathcal{E}_\g^{(2)}$ as a function of the potential depth $\left|v_0\right|$. The energy of a weakly bound ground state, that is $\abs{\E_\mathrm{g}^{(2)}} \ll 1$, shows a quadratic dependence on $\left|v_0\right|$ for different interaction potentials, illustrated for the case of a Gaussian-shaped potential $f_\textrm{G}$, Eq. \eqref{eq: def_Gaussian_potential}, and for a potential $f_\textrm{L}$  being characterized by the cube of a Lorentzian, Eq. \eqref{eq: def_Lorentzian3_potential}, by empty blue and filled red circles, respectively. The dashed red and blue lines are given by Eqs. (\ref{eq: Gaussian_potential_2_body_analytic}) and (\ref{eq: Lorentzian_potential_2_body_analytic}) accordingly, and confirm this dependency.}
\label{fig:universality_2_body}
\end{figure}

In order to reach sufficient convergence, we use $\mathcal{N} = 2500$ and numerically  obtain from Eq. (\ref{eq: two_body_discrete}) the potential depth $\abs{v_0}$ as a function of $\abs{\E_\g^{(2)}}$, displayed in Fig. \ref{fig:universality_2_body} by empty blue and filled red circles corresponding to $f_\mathrm{G}$ and $f_\mathrm{L}$.

In the limit $\left| v_0\right| \to 0$, the binding energy of the ground state is approximated by the expression~\cite{SIMON1976279,Berger2008}
\begin{equation}
\mathcal{E}_\g^{(2)} \cong -\frac{1}{2} v_0^2 \left[\int \mathrm{d}x\, f(x)\right]^2.
\label{eq: energy_approximation}
\end{equation}

From Eq.~\eqref{eq: energy_approximation} we obtain for our two test potentials the approximation
\begin{equation}
\left|\mathcal{E}_{\g}^{(2)}\right|\cong \frac{\pi}{2} v_0^2
\label{eq: Gaussian_potential_2_body_analytic}
\end{equation}
for $f=f_\textrm{G}$, and
\begin{equation}
\left|\mathcal{E}_{\g}^{(2)}\right|\cong \frac{9}{128}\pi^2 v_0^2
\label{eq: Lorentzian_potential_2_body_analytic}
\end{equation}
for $f=f_\textrm{L}$, depicted by a dashed blue and red line in Fig. \ref{fig:universality_2_body}, accordingly.

In the case of a contact interaction with $f=f_\delta$, Eq. \eqref{eq: def_delta_potential}, the relation given by Eq. \eqref{eq: energy_approximation} not only provides an approximation, but is exact as presented in Eq. (\ref{e2bcontact}). The pure quadratic dependence of the two-body binding energy on the potential depth $v_0$, and the fact that the contact interaction gives rise to only a single bound state for any value $v_0<0$ confirms that the corresponding two-body system is exactly on resonance. In the next sections we show that this unique feature has important consequences for the respective three-body system. 

\subsection{Universal limit}

\begin{figure*}[ht]
\begin{center}
\includegraphics[width=\textwidth]{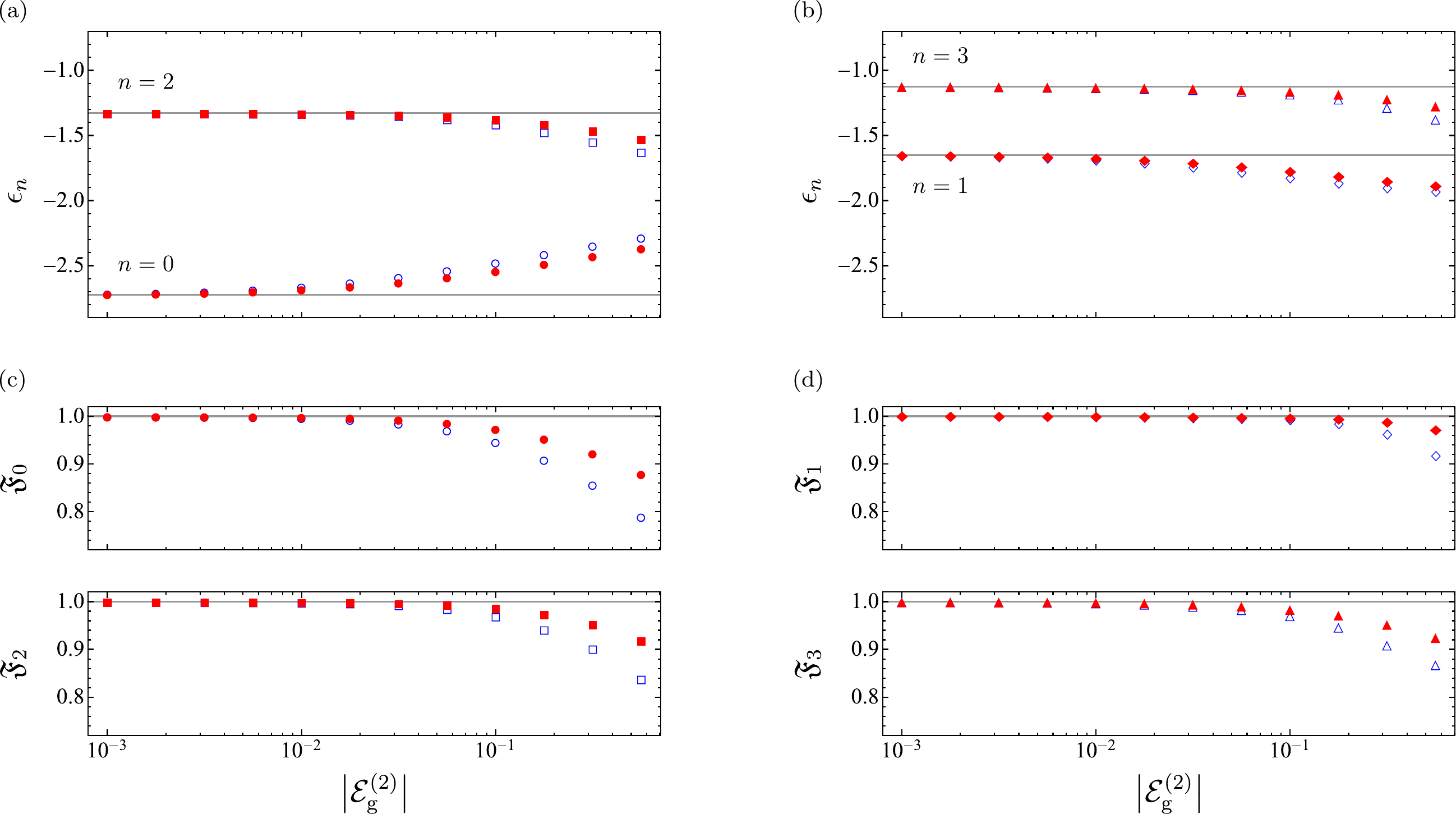}
\end{center}
\caption{Universal behavior of the 1D three-body system illustrated by the scaled three-body binding energy $\epsilon_n$, Eq. (\ref{epsilon}) (top), and the fidelity $\mathfrak{F}_n$, Eq. (\ref{eq: def_fidelity_general}) (middle and bottom), as a function of the two-body binding energy $\mathcal{E}_\g^{(2)}$  for three different interaction potentials: $f_\mathrm{G}$, Eq. (\ref{eq: def_Gaussian_potential}), (empty blue symbols), $f_\mathrm{L}$, Eq. (\ref{eq: def_Lorentzian3_potential}), (filled red symbols), and $f_\mathrm{\delta}$, Eq. (\ref{eq: def_delta_potential}), (gray lines). We display the first four energies (top)
and fidelities (middle and bottom) for (a), (c) \textit{bosonic} ($n=0,2$) and (b), (d) \textit{fermionic} ($n=1,3$) heavy particles for the mass ratio $M / m =20$. Close to a resonance of the heavy-light system, that is for $\mathcal{E}_\g^{(2)}\to 0$, each energy $\epsilon_n$ and fidelity $\mathfrak{F}_n$ with $n=0,1,2,3$ approaches the value determined by the contact interaction $f_\mathrm{\delta}$. Thus, we observe for all presented states a universal behavior, independent of the shape $f$ of the interaction potential.}
\label{fig:Three_body_universality}
\end{figure*}

We now consider the three-body problem in 1D with the heavy-light interaction potentials having the shape $f$ given by Eqs. (\ref{eq: def_Gaussian_potential}) and (\ref{eq: def_Lorentzian3_potential}), and compare the results to those obtained for the contact interaction, Eq. (\ref{eq: def_delta_potential}). In particular, we study the universal limit, that is $\mathcal{E}_\g^{(2)}\to 0$.

Following Appendix \ref{app:SM}, we apply a pseudospectral method based on the roots of the rational Chebyshev functions and represent the Schr\"odinger equation, Eq.  \eqref{TISGL}, as the eigenvalue problem
\begin{equation}
\left[-\frac{\alpha_x}{2} \boldsymbol{D_{xx}} -\frac{\alpha_y}{2} \boldsymbol{D_{yy}} + v_0 \left( \boldsymbol{F_+} + \boldsymbol{F_-}\right) \right] \vec{\psi} =  \mathcal{E} \vec{\psi}
\label{eq: three_body_discrete}
\end{equation}
for the eigenvalue $\mathcal{E}$ and the eigenvector $\vec{\psi}$ of size $\mathcal{M}$ approximating the three-body wave function $\psi(x,y)$. Here, the matrices $\boldsymbol{D_{xx}}$ and $\boldsymbol{D_{yy}}$ of size $\mathcal{M}\times \mathcal{M}$  correspond to the discretized second-order derivative with respect to $x$ and $y$, respectively. Moreover, the diagonal matrices $v_0 \boldsymbol{F_+}$ and $v_0 \boldsymbol{F_-}$ describe accordingly the interaction potentials along the lines $x+y/2$ and $x-y/2$, as shown in Appendix \ref{app: 2D_representation}.

We solve the finite-dimensional eigenvalue problem, Eq. \eqref{eq: three_body_discrete}, on the Data Vortex system DV206~\cite{DataVortex} by employing a parallelized version of the ARPACK software~\cite{ARPACK}, including an implementation of the Implicitly Restarted Arnoldi Method~\cite{IRAM}. In order to obtain sufficient convergence of all energies and the corresponding wave functions, we use a grid of size $\mathcal{M}=\mathcal{N}_x\cdot \mathcal{N}_y$ with $\mathcal{N}_x=512$ and $\mathcal{N}_y=256$.

\subsubsection{Energy spectrum}

For the mass ratio $M /m = 20$ we present in Fig.~\ref{fig:Three_body_universality} (a) and (b) the scaled energies $\epsilon_n$, Eq. \eqref{epsilon}, of the first four three-body bound states ($n=0,1,2,3$), as a function of the two-body binding energy $\left|\mathcal{E}_\g^{(2)}\right|$. In particular, we display by empty blue and filled red symbols the results for a Gaussian-shaped potential $f_\mathrm{G}$, Eq. \eqref{eq: def_Gaussian_potential}, and for a cubic Lorentzian-shaped potential $f_\mathrm{L}$, Eq. \eqref{eq: def_Lorentzian3_potential}.  The values of $\epsilon_n$ corresponding to the contact interaction $f_\delta$, Eq. \eqref{eq: def_delta_potential}, are independent of $\left|\mathcal{E}_\g^{(2)}\right|$ and are shown by gray lines, which reflects the feature of this interaction potential to support an exact two-body resonance. 

In Fig.~\ref{fig:Three_body_universality} we separate the cases of \textit{bosonic} heavy particles (a) associated with $n=0,2$ and \textit{fermionic} heavy particles (b) represented by $n=1,3$. For the interaction potentials with the shapes $f_\mathrm{G}$ and $f_\mathrm{L}$, we have used the numerically obtained relation between the two-body binding energy $\mathcal{E}_\g^{(2)}$ and the potential depth $v_0$ presented in Fig. \ref{fig:universality_2_body}.

In the case that the two-body heavy-light subsystem is close to a resonance, that is in the limit $\mathcal{E}_\g^{(2)}\to 0$, we observe a universal behavior of the scaled energies $\epsilon_n$ for all presented three-body bound states, that is for $n=0,1,2,3$. Moreover, we point out that different states approach the universal regime in different ways. Indeed, Fig.~\ref{fig:Three_body_universality} (a) and (b) show clearly that the difference of the energy $\epsilon_n$ and the corresponding universal limit for a fixed value of the two-body binding energy $\mathcal{E}_\g^{(2)}$ is usually  smaller in the case of higher excited states. 

\subsubsection{Wave functions}

Now we are in the position to compare not only the energies of the three-body bound states for different interaction potentials, but also the corresponding wave functions. For this purpose, we use again the fidelity 
\begin{equation}
\mathfrak{F}_n\equiv \left[\iint \textrm{d}x\textrm{d}y \, \psi_n(x,y) \psi_{n}^{\star}(x,y)\right]^2
\label{eq: def_fidelity_general}
\end{equation}
as a measure of the spatial overlap between the wave function $\psi_n^{\star}$, Eq. \eqref{wfscaling}, obtained for the case of a contact interaction $f_\delta$, Eq. \eqref{eq: def_delta_potential}, and the three-body wave function $\psi_{n}$ obtained as solution of Eq. \eqref{eq: three_body_discrete} for the interaction potential with the shape $f_\mathrm{G}$ and $f_\mathrm{L}$, respectively.

Using the relation between the potential depth $v_0$ and the two-body binding energy $\mathcal{E}_\g^{(2)}$ presented in Fig. \ref{fig:universality_2_body} we show in Fig. \ref{fig:Three_body_universality} the fidelities $\mathfrak{F}_n$ for (c) \textit{bosonic} and (d) \textit{fermionic} heavy particles as a function of the two-body binding energy  $\mathcal{E}_\g^{(2)}$ by empty blue for $f_\mathrm{G}$, and filled red symbols for $f_\mathrm{L}$. The value $\mathfrak{F}_n=1$ obtained for a contact interaction for any value $\abs{\E_\mathrm{g}^{(2)}}$ is displayed by gray lines. Independent of the shape $f$ of the interaction potential, the fidelity $\mathfrak{F}_n$ for $n=0,1,2,3$  approaches unity as $\mathcal{E}_\g^{(2)}\to 0$, and thus describes a perfect overlap of $\psi_{n}$ and $\psi_n^{\star}$. 

As shown by Eq. (\ref{wfscaling}) the coordinates of the wave function $\psi_n^{\star}$ are simply \textit{rescaled} as $\mathcal{E}_\g^{(2)}$ is varied due to the scaling property, Eq. (\ref{rescale3}), for $f=f_\delta$. Thus, close to the two-body resonance, $\mathcal{E}_\g^{(2)}\to 0$, the same behavior is true for the wave function $\psi_{n}$, revealing again the universal limit. 

\subsubsection{Summary}
In summary, the universal behavior of the three-body system is shown to appear both in the scaled energies $\epsilon_n$, Eq. \eqref{epsilon}, as well as in the scaling of the corresponding wave functions $\psi_n$. Moreover, we emphasize that the universal behavior manifests itself for all presented three-body bound states.

\section{Proof of universality}\label{sec:proof}
In the preceding section we have explored the universal behavior of the three-body bound state energies and the corresponding wave functions for interaction potentials of Gaussian and cubic Lorentzian shape, with the two-body ground state energy $\E_\g^{(2)}$ approaching zero. In this limit, we now prove the universality of the 1D three-body system for an arbitrary mass ratio $M/m$ and \textit{any} short-range interaction potential.

For this purpose we consider a heavy-light interaction of shape $f$ and recall the three-body Schr\"odinger equation in integral form, given by Eq. (\ref{stm1}). Next, we perform the substitutions $Z \equiv \ell_\g R_+' = \ell_\g (X' + Y'/2)$ on the first summand and  $Z \equiv \ell_\g R_-' = \ell_\g (X' - Y'/2)$ on the second one, to arrive at
\begin{align}\label{u3}
\tilde{\psi}&(X,Y) = -4 \ell_\g \abs{v_0} \iint \dint X'\dint Z f(Z) \nonumber \\
&\times \left[G_\epsilon^{(2)}(X-X',Y-2Z/\ell_\g+2X') \tilde{\psi}(X',2Z/\ell_\g-2X') \right.   \nonumber \\
& + \left. G_\epsilon^{(2)}(X-X',Y+2Z/\ell_\g-2X') \tilde{\psi}(X',-2Z/\ell_\g+2X') \right].
\end{align}
Here, we have used $\ell_\g$ defined by Eq. (\ref{ell}). With the approximate expression, Eq. (\ref{eq: energy_approximation}), for the two-body binding energy, valid in the regime $\E_\g^{(2)}\ll 1$, we obtain
\begin{align}\label{u4}
&\tilde{\psi}(X,Y) \cong -\frac{4}{\int \dint Z f(Z)} \iint \dint X'\dint Z f(Z) \nonumber \\
&\times \left[G_\epsilon^{(2)}(X-X',Y-2Z/\ell_\g +2X') \tilde{\psi}(X',2Z/\ell_\g -2X') \right.   \nonumber \\
& + \left. G_\epsilon^{(2)}(X-X',Y+2Z/\ell_\g -2X') \tilde{\psi}(X',-2Z/\ell_\g +2X') \right].
\end{align}

In the limit $\E_\mathrm{g}^{(2)} \to 0$, that is $\ell_\g \to \infty$, $G_\epsilon^{(2)}$ and $\tilde{\psi}$ become independent of $Z$, and as a result any dependence on the potential shape $f$ cancels in Eq. (\ref{u4}). Indeed, we retrieve Eq. (\ref{stm3}) valid for the contact interaction, with the solutions $\epsilon_n = \epsilon_n^{\star}$ and $\tilde{\psi}_n(X,Y) = \tilde{\psi}_n^{\star}(X,Y)$, as considered in Section \ref{sec:STM}. We emphasize that this is a consequence of the fact that Eq. (\ref{eq: energy_approximation}) is exact for this particular interaction potential. 

As a result, these universal constants $\epsilon_n^{\star}=\epsilon_n^{\star}(M/m)$ depend only on the mass ratio and can be used to formulate the relation
\begin{equation}
\E_n \cong \frac{1}{2}\epsilon_n^{\star}v_0^2 \left[\int\dint x f(x)\right]^2
\end{equation}
for the three-body binding energies as a function of the two-body interaction, valid for $\abs{v_0} \to 0$.

Hence, we have shown explicitly that \textit{all} scaled energies $\epsilon_n$, as well as the wave functions $\tilde{\psi}_n$ coincide with the results for the contact interaction, for \textit{any} short-range heavy-light interaction potential of shape $f$, provided we approach the two-body resonance defined by $\E_\mathrm{g}^{(2)} \to 0$.

\section{Conclusion and outlook} \label{sec:summary}
In this article, we have presented a quantum mechanical treatment of a heavy-heavy-light system confined to 1D. For a \textit{zero-range} heavy-light interaction we have studied the three-body energy spectrum and the corresponding wave functions using two different methods: (i) the Born-Oppenheimer approximation, and (ii) the exact integral equations of Skorniakov and Ter-Martirosian. In addition, for \textit{finite-range} interactions, we have investigated the universal limit of the three-body energies and the corresponding wave functions when the ground state energy of the heavy-light subsystem approaches zero.

In particular, for the case of a contact interaction we have explored the accuracy of the BO approximation in a regime of experimentally feasible mass ratios and found that the error in the energy spectrum drops rapidly from around 20\% in case of equal masses to below 2\% for rather extreme mass ratios $M/m \cong 22.2$ like in  $^{133}\mathrm{Cs}$--$^{6}\mathrm{Li}$ mixtures \cite{Naidon2017}. In addition, the ground state energy presented in Ref.~\cite{Mehta2014} for $M/m = 1$ agrees with our result.

The approximate BO wave functions are very close to the exact ones, since for $M/m = 25$ the fidelity reaches values up to $0.999$. As a result, the use of the approximate BO wave functions to calculate the mean value of the total Hamiltonian has significantly improved the accuracy of the three-body binding energies.

Moreover, by applying a pseudospectral method based on the roots of rational Chebyshev functions we have obtained the three-body energies and wave functions for the short-range interaction potentials of Gaussian and cubic Lorentzian shape. When the ground state energy of the heavy-light potential approaches zero, the universal behavior is apparent for both potentials, that is each three-body binding energy converges to the limit value determined by the zero-range contact interaction. We have also compared the associated wave functions to the ones provided by the contact interaction and we found that they follow a universal scaling law when the two-body resonance is approached.

Finally, we have demonstrated the universality of \textit{all} three-body bound states for \textit{any} short-range interaction potential when the heavy-light ground state energy approaches zero. Here, we recover the results for the contact interaction obtained within the STM approach. Hence, the three-body bound states for an arbitrary short-range interaction on resonance can be obtained by using a zero-range potential and applying the BO approximation, provided the mass ratio is sufficiently large. For experimentally relevant mass ratios, we present in Tab. \ref{table1} the universal constants determining the three-body binding energies in case of weak interactions.

\begin{table}
\caption{\label{table1}Universal constants $\epsilon_n^\star$ for different mass ratios commonly used in experiments performed with ultracold atoms.}
\begin{ruledtabular}
\begin{tabular}{l|ccc}  
& \multicolumn{3}{c}{atomic mixture $(M/m)$} \\
    & $^{87}\mathrm{Rb}$--$^{40}\mathrm{K}$ (2.2) & $^{87}\mathrm{Rb}$--$^{7}\mathrm{Li}$ (12.4) & $^{133}\mathrm{Cs}$--$^{6}\mathrm{Li}$ (22.2) \\
\hline
$\epsilon_0^\star$ \ \ & -2.1966	& -2.5963 & -2.7515  \\
$\epsilon_1^\star$ & -1.0520	& -1.4818 & -1.6904  \\
$\epsilon_2^\star$ & - 		& -1.1970 & -1.3604  \\
$\epsilon_3^\star$ & - 		& -1.0377 & -1.1479  \\
$\epsilon_4^\star$ & - 		& -1.0002 & -1.0525  \\
$\epsilon_5^\star$ & - 		& - 	  & -1.0040  \\
\end{tabular}
\end{ruledtabular}
\end{table}

We conclude by raising a few interesting generalizations of our approach. A nearly resonant excited state in the two-body system might lead to different features compared to the ones induced by the two-body ground state. According to Ref. \cite{Barlette2000} the two-body scattering in 1D only depends on the symmetry of the state. Based on this argument one might conclude that universal behavior in the three-body system only depends on the symmetry of the underlying two-body resonance. However, a rigorous study of this case is necessary to arrive at a definite statement. Moreover, we emphasize that further features might appear within a 3D consideration for the quasi-1D three-body system. Needless to say these questions go beyond the scope of the present article but will be addressed in a future publication.

\acknowledgments
We thank N.L. Harshman for fruitful discussions. Moreover, we gratefully acknowledge the support of Data Vortex Technologies which provided the Data Vortex system used for the numerical calculations. L.H. and M.A.E. thank the Center for Integrated Quantum Science and Technology (IQ$^{\rm ST}$) for financial support. This work is funded in part by the German-Israeli Project Cooperation DIP (Project No. AR 924/1-1, DU 1086/2-1). W.P.S.  is  most  grateful  to  Texas A$\&$M  University  for  a  Faculty  Fellowship  at  the  Hagler  Institute  for Advanced Study at the Texas A$\&$M University as well as to Texas A$\&$M AgriLife Research. The research of the IQ$^{\rm ST}$ is financially supported by the Ministry of Science, Research and Arts Baden-W\"urttemberg.

\appendix
\section{Born-Oppenheimer approximation for the three-body problem}\label{app:BO}
In this appendix we recall the main ideas of applying \cite{Efremov2009,Fonseca1979} the BO approximation to the three-body problem in the presence of a contact interaction. In particular, we derive the relevant formulas in zero order as well as the diagonal correction to the energy spectrum.

In the BO approach we represent the total wave function as the product
\begin{equation}
\tilde{\psi} \equiv \varphi(X,Y) \phi(Y),
\end{equation}
where we assign $\varphi$ and $\phi$ to the dynamics of the light and heavy particles, respectively.

When we apply the complete three-body Hamiltonian
\begin{equation}
\hat{H} \equiv \hat{H}_0 - \alpha_y \ddx{Y},
\end{equation}
with
\begin{equation}\label{hanull}
\hat{H}_0 \equiv -\alpha_x \ddx{X} - 2 \left[\delta(X+Y/2) + \delta(X-Y/2)\right],
\end{equation}
onto $\tilde{\psi}$, we obtain
\begin{align}\label{bofo1}
\hat{H}\tilde{\psi} =& \hat{H}_0\varphi \phi - \alpha_y\varphi \ddx{Y}\phi \nonumber \\
& -\alpha_y\left[\left(\ddx{Y} \varphi \right) + 2 \left(\dx{Y}\varphi\right)\dx{Y}\right]\phi.
\end{align}
So far our calculation is exact.

\subsection{Zero-order consideration}\label{app:BOZO}
In the zero-order approximation we neglect all derivatives of $\varphi$ with respect to the relative coordinate $Y$ of the heavy particles. This fact is emphasized by the vertical bar in the notation $\varphi(X,Y) \to \varphi(X|Y)$ and suggests to choose $\left\{\varphi \right\}$ as eigenbasis of $\hat{H}_0$, summarized by the light-particle Schr\"odinger equation
\begin{equation}\label{lpeqH0}
\hat{H}_0 \varphi = u \varphi.
\end{equation}

We obtain the BO potential $u=u(Y)$ by rewriting Eq. (\ref{lpeqH0}) with $\hat{H}_0$ given by Eq. (\ref{hanull}) in integral form
\begin{align} \label{bo2}
\varphi(X|Y) = -2 &\int \dint X'\,  G_u^{(1)}(X-X') \varphi(X'|Y) \nonumber \\
& \times \left[\delta\left(X'+Y/2\right) + \delta\left(X'-Y/2\right)\right] ,
\end{align}
where
\begin{equation}
G_u^{(1)}(X) \equiv -\frac{1}{2\sqrt{\abs{u}\alpha_x}} \e^{-\sqrt{\abs{u}/\alpha_x}\abs{X}}
\end{equation}
is the Green function of the one-dimensional free-particle Schr\"odinger equation for $u<0$.

Due to the delta functions, the integration over $X'$ can be performed immediately. By evaluating both sides of Eq. (\ref{bo2}) at the points $X = \pm Y/2$, we arrive at the transcendental equations
\begin{equation}\label{BOpoteq}
\e^{-\sqrt{\abs{u}/\alpha_x} \abs{Y}} = \pm \left(\sqrt{\abs{u}\alpha_x} - 1 \right)
\end{equation}
for the BO potentials $u_\pm = u_\pm (Y)$ with the solutions
\begin{equation}\label{BOpot_A}
u_\pm(Y) = -\frac{1}{\alpha_x}\left[\frac{\alpha_x}{\abs{Y}} W_0\left(\pm\frac{\abs{Y}}{\alpha_x}\e^{-\abs{Y}/\alpha_x}\right) +1 \right]^2
\end{equation}
in terms of the Lambert function $W_0$ \cite{abramowitz}, and the corresponding wave functions
\begin{equation}\label{lpwf_A}
\varphi_\pm(X|Y) = N_\pm\left[ \e^{-\sqrt{\abs{u_\pm}/\alpha_x}\abs{R_-}} \pm \e^{-\sqrt{\abs{u_\pm}/\alpha_x}\abs{R_+}}\right],
\end{equation}
where
\begin{equation}
N_\pm = \frac{1}{\sqrt{2}} \abs{ \sqrt{\frac{\alpha_x}{\abs{u_\pm}}} \pm \e^{-\sqrt{\abs{u_\pm}/\alpha_x} \abs{Y} }\left(\sqrt{\frac{\alpha_x}{\abs{u_\pm}}} + \abs{Y}\right) }^{-\frac{1}{2}}
\end{equation}
is a normalization factor.

The wave function $\phi_\pm = \phi_\pm(Y)$ is then a solution of the Schr\"odinger equation
\begin{equation}\label{hpeq_A}
\left[ - \alpha_y \ddx{Y} + u_\pm(Y) \right] \phi_\pm = \epsilon^{(\BO)} \phi_\pm,
\end{equation}
where the potential $u_\pm$ is given by Eq. (\ref{BOpot_A}), and $\epsilon^{(\BO)}$ denotes the three-body binding energy in the zero-order BO approximation.

\subsection{Diagonal correction to the energy spectrum}\label{app:BOfirstorder}
In the zero-order BO approximation, we neglect the last two terms in Eq. (\ref{bofo1}). However, in order to find corrections to these zero-order expressions, we have to consider now \textit{all} terms in Eq. (\ref{bofo1}). In this section, we derive the diagonal correction to the BO binding energies and find the connection to the mean value $\bar{\epsilon}^{(\BO)}$ defined in Eq. (\ref{ebo_1})

To distinguish between the so-called diagonal and non-diagonal contributions, we consider Eq. (\ref{bofo1}) with the zero-order solutions $\varphi = \varphi_i$, $\phi = \phi_{i n}$, where the subscript $i$ labels the light-particle channels, and $n$ numbers the state in each channel. We then  multiply Eq. (\ref{bofo1}) by $\varphi^{\ast}_k=\varphi_k$ from the left-hand side, perform the integration over $X$, and use the orthonormality of the light-particle states $\varphi_i$  ($i,k=\pm$) to write
\begin{align}\label{bofo2}
\int \dint X\, \varphi_k \hat{H} \varphi_i \phi_{in} =  \epsilon_n^{(\BO)} \delta_{ik} \phi_{in} + \delta\hat{H}_{ik}\phi_{in},
\end{align}
where $\delta_{ik}$ is the Kronecker delta and
\begin{align}\label{bofo3}
\delta\hat{H}_{ik} \equiv -\alpha_y \int \dint X\,\varphi_k \ddx{Y} \varphi_i - 2\alpha_y\left[ \int \dint X\,\varphi_k \dx{Y} \varphi_i\right] \dx{Y}.
\end{align}
Here we have used Eqs. (\ref{lpeqH0}) and (\ref{hpeq_A}) in order to identify the zero-order contribution $\epsilon^{(\BO)}$. For $k=i$ ($k\neq i$) we speak of the diagonal (non-diagonal) part.

In the diagonal case, the expression for $\delta H_{ii}$ simplifies, as the second term in Eq. (\ref{bofo3}) vanishes
\begin{equation}\label{bofo4}
\int \dint X \varphi_i \dx{Y} \varphi_i = \frac{1}{2}\dx{Y}\int \dint X\abs{\varphi_i}^2 = 0
\end{equation}
due to normalization.

In order to calculate the mean value
\begin{equation}\label{bofo4b}
\bar{\epsilon}_n^{(\BO)} \equiv \iint \dint X \dint Y\, \tilde{\psi}_n^{(\BO)}\hat{H}\tilde{\psi}_n^{(\BO)},
\end{equation}
defined by Eq. (\ref{ebo_1}), we multiply Eq. (\ref{bofo1}) by $\tilde{\psi}_{n}^{(\BO)} \equiv \varphi_+ \phi_{+ n}$ from the left, integrate over $X$, and use Eq. (\ref{bofo2}) for $k=i$ and Eq. (\ref{bofo4}), to arrive at
\begin{align}\label{bofo5}
\bar{\epsilon}_n^{(\BO)} =  \epsilon_n^{(\BO)} -\alpha_y \int \dint Y \abs{\phi_{+ n}}^2 \int \dint X \varphi_+ \ddx{Y}\varphi_+.
\end{align}

For the mean value, Eq. (\ref{bofo4b}), only the diagonal term ($k=i$) of Eq. (\ref{bofo3}) contributes to the correction. Hence, the expression $\bar{\epsilon}_n^{(\BO)}$ given by Eq. (\ref{bofo5}) equals the corrected BO binding energies, if couplings between different states are neglected, that is if $\delta \hat{H}_{ik} = 0$ for $k \neq i$.

\section{Pseudospectral methods}\label{app:SM}

Pseudospectral methods \cite{boyd2001chebyshev,trefethen2000spectral,BAYE20151} are an efficient tool to obtain a numerical solution of an ordinary or partial differential equation. In the following, we focus only on linear equations, where we represent the differential operators by matrices, and the unknown eigenfunctions by vectors. The corresponding eigenvalue problem of finite size can then be solved numerically.

Indeed, a key advantage of pseudospectral methods is the exponential convergence of the approximate solution to the exact one as the matrix size increases. For problems on a \textit{finite} domain, the convergence rate is usually geometric, whereas convergence for problems on an \textit{infinite} domain \cite{boyd2001chebyshev} is usually subgeometric. However, for a given matrix size the accuracy of the approximate solution is crucially determined by the deployed set of basis functions. Throughout this article we follow the suggestion of Boyd \cite{boyd2001chebyshev,BOYD1987112} and choose the rational Chebyshev functions as a basis.

In Appendix \ref{app: 1D_representation} we present the matrices used for a finite dimensional representation of a linear \textit{ordinary} differential equation of second order. A generalization of these matrices is obtained in Appendix \ref{app: 2D_representation} for the case of a linear \textit{partial} differential equation depending on two variables. Finally, we consider in Appendix \ref{app: eigenvalue_problem} the discretization of the eigenvalue problems analyzed in this article. 

\subsection{Matrix representation of 1D-problems}
\label{app: 1D_representation}

We begin by reviewing matrix representations of differential operators  defined on the finite domain $(-1,1)$, where Chebyshev polynomials \cite{boyd2001chebyshev} are used as basis functions. Next, we apply an algebraic map \cite{BOYD1987112} and obtain a finite dimensional representation of these operators on the complete real domain. 

\subsubsection{Finite domain}

First, we consider a grid based on the roots
\begin{equation}
\eta_i\equiv \cos\left[\frac{(2i+1)\pi}{2\mathcal{N}}\right]
\label{eq: Cheb_grid_points}
\end{equation} 
of the Chebyshev polynomial $T_{\mathcal{N}}=T_{\mathcal{N}}(\eta)$ of the first kind with degree $\mathcal{N}$ and $i=0,\ldots,\mathcal{N}-1$. 

This polynomial is defined by the recurrence relation
\begin{equation}
T_{\mathcal{N}}(\eta)\equiv 2\eta T_{\mathcal{N}-1}(\eta)-T_{\mathcal{N}-2}(\eta)
\label{eq: Cheb_polynomial_def}
\end{equation}
for $\mathcal{N}>1$ with $T_0(\eta)\equiv 1$ and $T_1(\eta)\equiv \eta$, where the argument $\eta$ is restricted to the finite interval $(-1,1)$.

For this grid the first-order derivative is represented \cite{boyd2001chebyshev} by the matrix
\begin{equation}
\left(\boldsymbol{\delta_1}\right)_{i,j}\equiv \begin{cases}
\frac{1}{2}\frac{\eta_i}{1-\eta_i^2}\,,&i=j\;,\\
\frac{(-1)^{(i+j)}}{\eta_i-\eta_j}\sqrt{\frac{1-\eta_j^2}{1-\eta_i^2}}\,,&i\neq j
\end{cases}
\end{equation}
with $i,j= 0,1,\ldots,\mathcal{N}-1$.

Similarly, the matrix representation of the second-order derivative reads
\begin{equation}
\left(\boldsymbol{\delta_2}\right)_{i,j}\equiv\begin{cases}
\frac{\eta_i^2}{(1-\eta_i^2)^2}-\frac{\mathcal{N}^2-1}{3(1-\eta_i^2)}\,,&i=j\,,\\
\left(\boldsymbol{\delta_1}\right)_{i,j}\left(\frac{\eta_i}{1-\eta_i^2}-\frac{2}{\eta_i-\eta_j}\right)\,,&i\neq j\,.
\end{cases}
\end{equation}

\subsubsection{Infinite domain}
Next, we consider the variable $x\in(-\infty, \infty)$ and extend the previous grid to an infinite domain. For this purpose we introduce the new grid points
\begin{equation}
x_i\equiv\frac{L\eta_i}{\sqrt{1-\eta_i^2}}
\label{eq: Rat_Cheb_grid_points}
\end{equation}
obtained from the old ones $\eta_i$ given by Eq.~\eqref{eq: Cheb_grid_points} by applying an algebraic map~\cite{BOYD1987112,boyd2001chebyshev}. The mapping parameter~$L$ determines the effective size of the grid. 

The grid points $x_i$ are the roots of the rational Chebyshev functions
\begin{equation}
TB_{\mathcal{N}}(x)\equiv T_{\mathcal{N}}\left(\frac{x}{\sqrt{L^2+x^2}}\right)
\end{equation}
given in terms of the Chebyshev polynomials $T_{\mathcal{N}}$ of degree $\mathcal{N}$ as defined by Eq.~\eqref{eq: Cheb_polynomial_def}.

As a result, the discrete representation of the differential operator $\frac{\textrm{d}}{\textrm{d}x}$ is given by the matrix
\begin{equation}
\boldsymbol{D_1}\equiv  \boldsymbol{A}\cdot \boldsymbol{\delta_1}
\end{equation}
of size $\mathcal{N}\times \mathcal{N}$ where the elements of the diagonal matrix~$\boldsymbol{A}$ read
\begin{equation}
(\boldsymbol{A})_{i,i}\equiv \frac{1}{L}\left(1-\eta_i^2\right)^{\frac{3}{2}}\,.
\end{equation}

Similarly, the discrete representation 
\begin{equation}
\boldsymbol{D_2}\equiv \boldsymbol{A}^2 \cdot \boldsymbol{\delta_2}+\boldsymbol{B} \cdot \boldsymbol{\delta_1},
\label{eq: D2_diff_matrix}
\end{equation}
of the second-order differential operator $\frac{\textrm{d}^2}{\textrm{d}x^2}$ is also determined by the diagonal matrix $\boldsymbol{B}$ with elements
\begin{equation}
(\boldsymbol{B})_{i,i}\equiv -\frac{3}{L^2}\eta_i\left(1-\eta_i^2\right)^2.
\end{equation}

Additionally, a matrix representation of any function $f=f(x)$ is given by the diagonal matrix $\boldsymbol{F}$ of size $\mathcal{N}\times \mathcal{N}$ with the elements
\begin{equation}
(\boldsymbol{F})_{i,i}\equiv f(x_i)\,.
\label{eq: 1D_potential_matrix}
\end{equation}
obtained by evaluating $f$ at the grid points $x_i$.

\subsection{Matrix representation of 2D-problems}
\label{app: 2D_representation}
Now, we generalize our grid to accommodate a \textit{partial} differential equation depending on the two independent variables $x$ and $y$. 

We introduce the grid points $(x_i, y_j)$ with
\begin{equation}
x_i\equiv  L_x \eta_{x,i} \left(1-\eta_{x,i}^2\right)^{-1/2} 
\label{eq: 2D_gridpoints_x}
\end{equation}
and
\begin{equation}
y_j\equiv L_y \eta_{y,j} \left(1-\eta_{y,j}^2\right)^{-1/2} 
\label{eq: 2D_gridpoints_y}
\end{equation}
where $\eta_{x,i}\equiv \eta_i$ for $i=0,1,\ldots, \mathcal{N}_x-1$ and $\eta_{y,j}\equiv \eta_{j}$ for $j=0,1,\ldots,\mathcal{N}_y-1$. The grid points $\eta_i$ and $\eta_j$ are defined by Eq. \eqref{eq: Cheb_grid_points} and the integers $\mathcal{N}_x$ and $\mathcal{N}_y$ denote the number of grid points used for the variables $x$ and $y$, respectively, with the corresponding mapping parameters $L_x$ and $L_y$.

The discrete representation of the partial second-order derivative $\frac{\partial^2}{\partial x^2}$ reads
\begin{equation}
\boldsymbol{D_{xx}}=\boldsymbol{D_{2,x}}\otimes \boldsymbol{\mathds{1}_y}\,.
\label{eq: 2D_Dxx_matrix}
\end{equation}
Here, the matrix $\boldsymbol{D_{2,x}}\equiv \boldsymbol{D_2}$ of the size $\mathcal{N}_x\times \mathcal{N}_x$ is given by Eq. \eqref{eq: D2_diff_matrix}, whereas $\boldsymbol{\mathds{1}_y}$ denotes the identity matrix of size $\mathcal{N}_y\times \mathcal{N}_y$. Thus, the matrix $\boldsymbol{D_{xx}}$ has the size $\mathcal{M}\times \mathcal{M}$ with $\mathcal{M}\equiv\mathcal{N}_x\cdot \mathcal{N}_y$. 

In a similar way, the partial second-order derivative $\frac{\partial^2}{\partial y^2}$ is represented by the $\mathcal{M}\times \mathcal{M}$ matrix
\begin{equation}
\boldsymbol{D_{yy}}=\boldsymbol{\mathds{1}_x}\otimes \boldsymbol{D_{2,y}}\,,
\label{eq: 2D_Dyy_matrix}
\end{equation}
where $\boldsymbol{\mathds{1}_x}$ is the identity matrix of size $\mathcal{N}_x\times \mathcal{N}_x$, and the matrix $\boldsymbol{D_{2,y}}\equiv \boldsymbol{D_2}$ of the size $\mathcal{N}_y\times \mathcal{N}_y$ is given by Eq. \eqref{eq: D2_diff_matrix}. 

Moreover, similar to Eq. \eqref{eq: 1D_potential_matrix}, a function depending on the variables $x$ and $y$ is represented by a diagonal matrix. In particular, the function $f(x\pm y/2)$ is given by the diagonal matrix $\boldsymbol{F_\pm}$ of size $\mathcal{M}\times \mathcal{M}$ with the elements
\begin{equation}
(\boldsymbol{F_\pm})_{i \mathcal{N}_y+j, i\mathcal{N}_y+j}\equiv f\left(x_i\pm y_j/2 \right)\,.
\label{eq: 2D_potential_matrix}
\end{equation}

\subsection{Eigenvalue problem}
\label{app: eigenvalue_problem}
Finally, we possess all ingredients to represent the stationary Schr\"odinger equation for the two-body, and the three-body system as an eigenvalue problem in terms of matrices provided by a pseudospectral method being determined by the roots of rational Chebyshev functions. 

We start by discussing the two-body system described the 1D Schr\"odinger equation
\begin{equation}
\left[-\frac{1}{2}\frac{\textrm{d}^2}{\textrm{d}x^2}+v_0 f(x)\right]\psi^{(2)}(x)=\mathcal{E}^{(2)}\psi^{(2)}(x)\,
\end{equation}
given by Eq. \eqref{eq: Schroedinger_2body_dmless} where we have used the definition, Eq. \eqref{interaction}, of the interaction potential.

Using the matrices $\boldsymbol{D_2}$ and $\boldsymbol{F}$ of size $\mathcal{N}\times \mathcal{N}$ defined by Eqs. \eqref{eq: D2_diff_matrix} and \eqref{eq: 1D_potential_matrix}, we arrive at the eigenvalue problem
\begin{equation}
\left[-\frac{1}{2} \boldsymbol{D_2}+v_0 \boldsymbol{F}\right]\vec{\psi}^{(2)}=\mathcal{E}^{(2)}\vec{\psi}^{(2)}
\end{equation}
for the eigenvector
\begin{equation}
\vec{\psi}^{(2)}\equiv \left\{\psi^{(2)}(x_0),\psi^{(2)}(x_1),\ldots, \psi^{(2)}(x_{\mathcal{N}-1})\right\}^{\textrm{T}}\,,
\end{equation}
where the grid points $x_0,\ldots,x_{\mathcal{N}-1}$ are given by Eqs. \eqref{eq: Cheb_grid_points} and \eqref{eq: Rat_Cheb_grid_points}. 

Next, we consider the three-body system governed by the Schr\"odinger equation 
\begin{equation}
\left\{-\frac{\alpha_x}{2}\frac{\partial^2}{\partial x^2}-\frac{\alpha_y}{2}\frac{\partial^2}{\partial y^2}+v_0\left[f(r_+)+f(r_-)\right]\right\}\psi=\mathcal{E}\psi
\label{eq: app_Schroedinger_2D}
\end{equation}
given by Eq. \eqref{TISGL} with $r_\pm \equiv x\pm y/2$ where $\alpha_x$ and $\alpha_y$ are defined by Eqs. \eqref{eq: alphax} and \eqref{eq: alphay}, respectively.

Using Eqs. \eqref{eq: 2D_Dxx_matrix}, \eqref{eq: 2D_Dyy_matrix} and \eqref{eq: 2D_potential_matrix} for the matrices $\boldsymbol{D_{xx}}$, $\boldsymbol{D_{yy}}$ and $\boldsymbol{F_\pm}$ of size $\mathcal{M}\times \mathcal{M}$, we obtain the eigenvalue problem 
\begin{equation}
\left[-\frac{\alpha_x}{2} \boldsymbol{D_{xx}}-\frac{\alpha_y}{2} \boldsymbol{D_{yy}}+v_0\left(\boldsymbol{F_+} + \boldsymbol{F_-}\right)\right]\vec{\psi}=\mathcal{E}\vec{\psi}
\label{eq: app_Schroedinger_2D_matrix}
\end{equation}
for the eigenvector 
\begin{align}
\vec{\psi}\equiv &\left\{\psi\left(x_0,y_0\right), \psi\left(x_0,y_1\right),\ldots, \psi\left(x_0,y_{\mathcal{N}_y-1}\right),\right. \nonumber\\
&\left. \psi\left(x_1,y_0\right),\ldots, \psi\left(x_{\mathcal{N}_x-1},y_{\mathcal{N}_y-1}\right)\right\}^\textrm{T}
\end{align}
determining the values of $\psi(x,y)$ at the grid points $(x_i,y_j)$ given by Eqs. \eqref{eq: 2D_gridpoints_x} and \eqref{eq: 2D_gridpoints_y}. 

We emphasize that the size of the matrices used in Eq. \eqref{eq: app_Schroedinger_2D_matrix} reduces by a factor of four when we take advantage of the symmetries of Eq. \eqref{eq: app_Schroedinger_2D} with respect to the transformations $x\rightarrow -x$ and $y\rightarrow -y$.
In our calculations we have made use of these symmetries and modified a method \cite{FOR95,trefethen2000spectral} originally suggested to improve pseudospectral grids for polar and spherical geometries. In this way we could reduce the size of our matrices while keeping the same accuracy.

\bibliography{fewbody}

\end{document}